\documentclass[letterpaper]{article} 
\usepackage{aaai2027}
\usepackage{times,helvet,courier,graphicx,natbib,caption,amsmath,amssymb,amsfonts,booktabs,algorithm,algpseudocode,comment,array,booktabs,multirow,tabularx,xcolor,pifont,dblfloatfix}
\usepackage[hyphens]{url} 
\usepackage[misc]{ifsym}
\usepackage{tcolorbox}
\tcbuselibrary{breakable,listings,skins}

\usepackage{booktabs}
\usepackage{multirow}
\usepackage{pdflscape}
\usepackage{adjustbox}
\usepackage{caption}
\usepackage{changepage}
\usepackage{tcolorbox}
\tcbuselibrary{breakable,listings}
\definecolor{tmplTitleBg}{HTML}{3B3B3B}
\definecolor{tmplTitleFg}{HTML}{FFFFFF}
\definecolor{tmplBodyBg}{HTML}{F3F3F3}
\definecolor{tmplFrame}{HTML}{2B2B2B}
\definecolor{phUser}{HTML}{3B6EEA}   
\definecolor{phGoal}{HTML}{19A35B}   
\definecolor{phTools}{HTML}{B65AD4}  
\definecolor{myblue}{HTML}{0072B2}
\definecolor{shepherd}{HTML}{0072B2}

\tcbset{
  tmplbox/.style={
    enhanced,
    breakable,
    colback=tmplBodyBg,
    colframe=tmplFrame,
    boxrule=0.7pt,
    arc=2.2mm,
    left=7pt,right=7pt,top=6pt,bottom=6pt,
    title={#1},
    coltitle=tmplTitleFg,
    colbacktitle=tmplTitleBg,
    fonttitle=\bfseries, 
    attach boxed title to top left={yshift=-1.2mm, xshift=2.0mm},
    boxed title style={sharp corners, boxrule=0pt, arc=2.2mm},
  }
}

\newtcolorbox{rqfinding}{
  colback=black!2,
  colframe=black!55,
  boxrule=0.45pt,
  arc=0.8mm,
  left=4pt,
  right=4pt,
  top=2.5pt,
  bottom=2.5pt,
  boxsep=0pt,
  before skip=4pt,
  after skip=5pt,
  fontupper=\small
}

\newcommand{\ActBench}{\textsc{ActBench}}
\newcommand{\ind}{\mathbb{I}}

\title{
\ActBench{}: Self-Evolving Benchmark of Behavioral Safety in Cowork Agents
}
\author{Hongwei Yao${^{1}}$\textsuperscript{$\dagger$}, Yiming Liu${^{2}}$\textsuperscript{$\dagger$}, Meihui Chen${^{4}}$, Jieling Chen${^{4}}$, \\
Zikun Chen${^{2}}$, Yiling He$^{^{3}\text{\Letter}}$, Wangze Ni${^{2}}$, Cong Wang${^{1}}$, Kui Ren${^{2}}$
\thanks{
\noindent 
\text{\Letter} Corresponding Author \\
${^{1}}$ City University of Hong Kong, Hong Kong \\
${^{2}}$ Zhejiang University, China \\
${^{3}}$ University College London, United Kingdom \\
${^{4}}$ Xiaomi Corporation, China}}

\affiliations{}
\begin{document}
\maketitle
\begin{abstract} 
Cowork agents may complete benign tasks while disclosing protected data, manipulating unauthorized state, invocate unauthorized API. We define \emph{behavioral safety} and introduce \ActBench{}, a self-evolving benchmark that evaluates such behavior risk from execution trajectories rather than final responses. Each case pairs a benign task with an adversarial variant that preserves its instruction, configuration, initial state, rating model, and trusted records while injecting a task-reachable payload. \ActBench{} contains 600 cases from 213 scenarios, spanning 15 risk behaviors, six execution spaces, and 48 web-service APIs.
To move beyond static payloads, we propose a reward-guided beam search method that jointly optimizes attack effectiveness and task utility, while reflection diagnoses failed execution checkpoint and guides payload revision. 
Besides, we propose a dual evidence verification mechanism that verifies agent execution safety and utility through log evidence and LLM-based trajectory evidence.
\textbf{We evaluate 15 LLMs and 6 open-source cowork agents over 24,000 trajectories}. 
Under a fixed harness, attack success rates ranges from 10.1\% to 94.4\% across models, while under a fixed base model, they range from 73.7\% to 94.4\% across agents.
These results show greater variation across models than agent harness, while attacks remain highly successful across all tested harnesses.
Our benchmark is released at: \url{https://github.com/zjuicsr/ActBench}.
\end{abstract}

\section{Introduction}
\label{sec:introduction}
Cowork agents increasingly operate across heterogeneous execution contexts, including workspace files, web content, persistent memory, external tools, and reusable skills. This broad access enables powerful automation but also creates risks that unfold across actions. 
OpenClaw~\citep{openclaw2026introducing} provides a prominent example: shortly after attracting widespread adoption in early 2026, the advisory for \textit{CVE-2026-25253} documented gateway-token exfiltration that could enable operator access and host code execution~\citep{openclaw2026cve}.
Later studies also reported credential disclosure, poisoned skills, and persistent state tampering~\citep{wang2026security,wang2026systematic,jiang2026humans,he2025artificial}. 
These incidents motivate a trajectory-based safety evaluation, which moves beyond response-level refusal metrics to characterize operational risk arising across sequences of actions and state transitions during agent execution.

To formalize this, we define \emph{behavioral safety} as whether an agent’s execution remains within the permissions and state changes required by a benign task. A risk behavior is identified when trusted records show that an out of scope action produced a prohibited effect through the observed execution trajectory. This definition separates safety from task completion, covering cases in which an agent may finish the requested task while disclosing protected data, modifying an unauthorized state, persisting an untrusted instruction, exceeding a resource limit, or claiming completion without the required effect. 
Behavioral safety must therefore be assessed from the execution evidence that determines what actions occurred and what effects they produced.

\paragraph{Motivation and Gaps.}
Recent benchmarks extend agent safety evaluation from response moderation to tool execution~\cite{jiang2026harmfulskillbench,yao2026red}. 
However, their coverage remains fragmented across distinct attack surfaces and evaluation settings, leaving three important gaps for stateful cowork agents.
\textbf{\textit{Gap 1: Existing data cover narrow and disjoint attack surfaces.}} AgentDojo provides 629 cases over untrusted tool outputs, WASP targets web content, and MCPTox evaluates 1,348 tool poisoning cases~\citep{debenedetti2024agentdojo,evtimov2025wasp,wang2026mcptox}. AgentPoison and MINJA isolate memory poisoning~\citep{chen2024agentpoison,dong2025minja}. No single benchmark jointly evaluates task artifacts, tool interfaces, persistent state, context pressure, and permission composition under task configurations.
\textbf{\textit{Gap 2: Existing evaluations lack causal attribution for behavioral safety.}} Outcome checks, attacker progress scores, and LLM judges do not jointly verify the realized effect, the provenance linked execution path, and the operation's configuration status~\citep{debenedetti2024agentdojo,evtimov2025wasp,ruan2024toolemu}. Consequently, a blocked call, a preexisting state, or an authorized transition may receive the same label as a realized policy violation. Such scores neither identify the failed enforcement point nor support category matched analysis.
\textbf{\textit{Gap 3: Static payload suites cannot adapt to execution feedback.}}
Fixed injection suites in existing work do not optimize payloads based on trajectory-level failure diagnoses. Therefore, a payload remains unchanged when an agent does not observe them, rejects the induced action, or fails before the target effect. 
This static construction largely limits attack coverage across models and harnesses.

\paragraph{Our Approach.}
We present \ActBench{}, a cowork agent behavioral safety benchmark with comprehensive test cases and reward-guided payload search.
To address Gap 1, \ActBench{} spans 15 risk behaviors, six execution spaces, 48 web service APIs, and 213 operational scenarios. Each malicious case injects one task reachable payload, while its benign counterpart preserves the instruction, task configuration, initialized state, utility grading criterion, attack grading criterion, and trusted logs.
To address Gap 2, we model cowork execution over six spaces and associate each behavior with a protected boundary, propagation path, failed enforcement condition, and evidence schema. Dual evidence verification triggers an attack criterion only when state evidence confirms the prohibited effect, validated event identifiers establish the observed path, and the task configuration excludes the operation. 
To address Gap 3, \ActBench{} employs a self-evolving optimization mechanism to find higher scoring payloads. Reward-guided beam search ranks edits by a geometric score over mean Attack Grading Score (\(\mathrm{AGS}\)) and Utility Grading Score (\(\mathrm{UGS}\)) across repeated rollouts. 
For unsuccessful attempts, reflection-based deep probing identifies the earliest failed checkpoint and revises the same payload field. 
Search and evaluation use disjoint rollouts to prevent optimization leakage. We further control comparison conditions by fixing the execution harness when comparing models and fixing the base model when comparing agent frameworks.

Our contributions are summarized as follows:
\begin{itemize}
\item We define \emph{behavioral safety} as whether an agent’s execution remains within the permissions and state changes required by a benign task. We operationalize 15 risk behaviors through trajectory predicates across four propagation families and six execution spaces, together with \(\mathrm{AGS}\) and \(\mathrm{UGS}\) for separate measurement of prohibited effects and required task effects.
\item We construct \ActBench{} with 300 benign and malicious pairs from 213 operational scenarios. The four-stage construction combines an empirical strategy pool, reward-guided beam search, reflection-based deep probing, and dual evidence verification. 
\item We evaluate 15 LLMs and 6 open-source cowork agents, including OpenClaw, Hermes, Claude Code, OpenAgent, OpenCode, and QwenPaw, across 20 controlled configurations and 24,000 execution trajectories. 
The results reveal substantial safety variation: \(\mathrm{ASR}\) ranges from 10.1\% to 94.4\% across LLMs under the same harness, and from 73.7\% to 94.4\% across agents under the same model.
\end{itemize}

\section{Related Work}
\label{sec:related}
\paragraph{Agent safety benchmarks.}
Existing agent safety benchmarks evaluate either compliance with harmful requests or security failures under adversarial execution conditions. 
For harmful-task evaluation, ToolEmu evaluates 36 high stakes toolkits in 144 cases and judges 68.8\% of detected failures plausible under deployment review~\citep{ruan2024toolemu}; AgentHarm contains 110 malicious tasks across 11 harm categories~\citep{andriushchenko2025agentharm}; RedCode includes 4,050 risky execution cases across 25 scenarios and 160 malicious code generation prompts~\citep{guo2024redcode}; and OS Harm extends harmful task evaluation to computer use agents~\citep{kuntz2025osharm}. 
For adversarial execution, AgentDojo provides 97 tasks and 629 security cases over untrusted tool returns~\citep{debenedetti2024agentdojo}; WASP observes partial attacker progress in up to 86\% of web injection cases~\citep{evtimov2025wasp}; MobileSafetyBench contains 200 daily scenario tasks and 50 indirect injection tasks in Android emulators~\citep{lee2026mobilesafety}; and Agent Security Bench evaluates prompt injection, memory poisoning, Plan of Thought backdoors, and mixed attacks, with a highest average attack success rate of 84.30\%~\citep{zhang2025asb}. These suites nevertheless isolate execution settings and omit matched configurations spanning artifacts, tool interfaces, persistent state, context pressure, and permission composition. \ActBench{} evaluates these surfaces through matched benign inputs and task configurations, with joint utility and safety outcomes across six execution spaces.

\paragraph{Prompt injection attacks.} 
Prompt injection attacks embed competing instructions within application data, causing a model or agent to treat untrusted content as executable instructions~\citep{liu2024promptinjection,shi2025quantifying}.
Initial work focuses on single-turn model settings~\citep{zverev2025separation}, while later studies extend injection to agent-accessible data resources~\citep{wang2026mcptox}.
For example, EIA attacks compromised webpages and achieves up to 70\% success in extracting specific personally identifiable information~\citep{liao2025eia}, AgentPoison assumes direct poisoning access to memory or knowledge bases~\citep{chen2024agentpoison}, whereas MINJA injects malicious records through query only interaction~\citep{dong2025minja}.

\paragraph{Safety detection and auditing.}
Existing work improves agent safety through attack detection, model alignment, vulnerability discovery, and automated auditing~\cite{he2026attriguard}.
DataSentinel develops game theoretic prompt injection detection~\citep{liu2025datasentinel}. SecAlign applies preference optimization and lowers the strongest tested attack to 8\% in its Llama 3 8B setting while retaining base model utility~\citep{chen2025secalign}. AgentFuzz tests 20 open source agents and finds 34 high risk zero day taint style vulnerabilities across 14 agents~\citep{liu2025agentfuzz}. AgentAuditor introduces ASSEBench with 2,293 annotated interactions across 15 risk types and 29 application scenarios~\citep{luo2025agentauditor}.

\section{Risk Behavior Taxonomy}
\label{sec:taxonomy}
\paragraph{Agent execution and test case.}
We model cowork agent across six execution spaces, including context \(\mathsf{C}\), reasoning \(\mathsf{R}\), safety policy \(\mathsf{P}\), action \(\mathsf{A}\), persistent memory \(\mathsf{M}\), and environment \(\mathsf{E}\). At execution step \(t\), the system state is \(y^{t}=(c^{t},r^{t},p^{t},a^{t},m^{t},e^{t}).\)
Starting from \(y^{t}\), the reasoning model returns the execution trajectory \(\tau=\pi_{\phi}(y^{t})\), which contains the sequence of model-visible context, reasoning, actions, tool results, memory events, and environment records.
A test case is \(x_i=(c,m,e,\{u_i,\mathcal{U}_i,\mathcal{A}_i\})\), where \(u_i\) is the benign user instruction. The sets \(\mathcal{U}_i\) and \(\mathcal{A}_i\) contain utility and attack grading criteria, respectively. Benchmark construction may edit \(c\), \(m\), or \(e\). It keeps \(u_i\), \(\mathcal{U}_i\), and \(\mathcal{A}_i\) fixed.

\begin{table}[t!]
\centering
\scriptsize
\setlength{\tabcolsep}{3pt}
\renewcommand{\arraystretch}{1.05}
\renewcommand{\tabularxcolumn}[1]{m{#1}}
\newcommand{\tworowstrut}{%
    \rule[-1\baselineskip]{0pt}{2.05\baselineskip}%
}
\begin{tabularx}{\linewidth}{
    >{\centering\arraybackslash}m{0.05\linewidth}
    >{\raggedright\arraybackslash}m{0.36\linewidth}
    >{\raggedright\arraybackslash}X
}
\specialrule{1pt}{2pt}{2pt}
\textbf{ID} & \textbf{Risk Behavior} & \textbf{Description} \\
\midrule

\(\text{B}_1\) & Instruction Injection
& \multirow[c]{4}{=}{%
    \raggedright
    \textbf{Family:} Context manipulation in $\mathsf{C}$\par
    \textbf{Effect:} Untrusted content alters objectives, disclosure
    decisions, or credential handling.\par
    \textbf{Path:}
    $\mathsf{C}_{\mathrm{adv}}
    \rightarrow \mathsf{R}
    \rightarrow \mathsf{A}
    \rightarrow \mathsf{E}$%
} \\
\(\text{B}_2\) & Goal Hijacking       & \\
\(\text{B}_3\) & Data Exfiltration    & \\
\(\text{B}_4\) & Credential Exposure  & \\
\midrule

\tworowstrut \(\text{B}_5\) & \tworowstrut Memory Poisoning
& \multirow[c]{2}{=}{%
    \raggedright
    \textbf{Family:} State integrity in $\mathsf{M}$ and $\mathsf{E}$\par
    \textbf{Effect:} An unverified persistent record alters a later
    decision or action.\par
    \textbf{Path:}
    $\mathsf{M}_{\mathrm{adv}}/\mathsf{E}_{\mathrm{adv}}
    \rightsquigarrow \mathsf{C}
    \rightarrow \mathsf{R}
    \rightarrow \mathsf{A}$
} \\
\tworowstrut \(\text{B}_6\) & \tworowstrut State Tampering & \\
\midrule

\(\text{B}_7\) & Deceptive Tool Invocation
& \multirow[c]{4}{=}{
    \raggedright
    \textbf{Family:} Action misuse in $\mathsf{A}$\par
    \textbf{Effect:} Tool selection or parameters exceed the operation,
    API, object.\par
    \textbf{Path:}
    $\mathsf{C}
    \rightarrow \mathsf{R}
    \rightarrow \mathsf{A}_{\mathrm{adv}}
    \rightarrow \mathsf{E}$%
} \\
\(\text{B}_8\)  & Unauthorized CMD Execution & \\
\(\text{B}_9\)  & Unauthorized API Invocation & \\
\(\text{B}_{10}\) & Tool Scope Escalation       & \\
\midrule

\(\text{B}_{11}\) & Resource Exhaustion
& \multirow[c]{5}{=}{%
    \raggedright
    \textbf{Family:} Policy mediation failure in $\mathsf{P}$\par
    \textbf{Effect:} Enforcement fails for resource limits, canonical
    execution, reporting, or context allocation.\par
    \textbf{Path:}
    $\mathsf{C}
    \rightarrow \mathsf{R}
    \rightarrow \mathsf{P}_{\mathrm{fail}}
    \rightarrow \mathsf{A}
    \rightarrow \mathsf{E}$%
} \\
\(\text{B}_{12}\) & Obfuscated Execution & \\
\(\text{B}_{13}\) & False Reporting      & \\
\(\text{B}_{14}\) & Context Flooding     & \\
\(\text{B}_{15}\) & Permission Chaining  & \\
\specialrule{1pt}{2pt}{2pt}
\end{tabularx}

\caption{Risk behavior taxonomy defined by where each behavior becomes operational and how its effect propagates through the trajectory path.}
\label{tab:taxonomy}
\end{table}

\paragraph{Risk taxonomy.}
\label{para:taxonomy}
We propose a trajectory-based risk behavior taxonomy. In contrast to taxonomies based on payload or final response, our taxonomy assigns a behavior through its realized propagation path in \(\tau\). We define 15 risk behaviors \(\text{B}_j\), where \(j\in\{1,\ldots,15\}\), and organize them into four families in Table~\ref{tab:taxonomy}. Context manipulation behaviors \(\text{B}_1\sim\text{B}_4\) arise when untrusted content changes the interpreted objective, disclosure decision, credential handling. State integrity behaviors \(\text{B}_5\sim\text{B}_6\) capture cases in which persistent memory or environment state alters a later decision. Action misuse behaviors \(\text{B}_7\sim\text{B}_{10}\) concern tool selection, command execution, API invocation, and argument scope that exceed the task authorization. Policy mediation failures \(\text{B}_{11}\sim\text{B}_{15}\) arise when the safety policy layer fails to enforce a resource bound, canonical execution rule, reporting requirement, context constraint, or composed permission.

\paragraph{Criterion-based scoring.}
For trajectory \(\tau\), each attack grading criterion \(a\in\mathcal{A}_i\) and each utility grading criterion \(u\in\mathcal{U}_i\) returns a binary value. We define
\begin{equation}
\resizebox{0.9\linewidth}{!}{$
\begin{aligned}
g_a(x_i)&=\frac{1}{|\mathcal{A}_i|}
\sum_{a\in\mathcal{A}_i}\ind[a(\tau)=1], \quad
g_u(x_i)&=\frac{1}{|\mathcal{U}_i|}
\sum_{u\in\mathcal{U}_i}\ind[u(\tau)=1].
\label{eq:criterion_score}
\end{aligned}
$}
\end{equation}
We call \(g_a(x_i)\) the attack grading score and \(g_u(x_i)\) the utility grading score. Both scores lie in \([0,1]\) and measure criterion coverage. A value of \(g_a(x_i)=1\) means that every attack criterion returns 1, and a value of \(g_u(x_i)=1\) means that every required task criterion returns 1. Intermediate values report the fraction of satisfied criteria rather than a probability estimate. A rollout is task complete when \(g_u(x_i)=1\), while full attack success requires \(g_a(x_i)=1\). We propose a dual evidence verification mechanism that validates the final score through log-based and LLM-based trajectory evidence.

\section{Benchmark Construction}
\label{sec:construction}
\begin{figure*}[t!]
\centering
\includegraphics[width=0.95\linewidth]{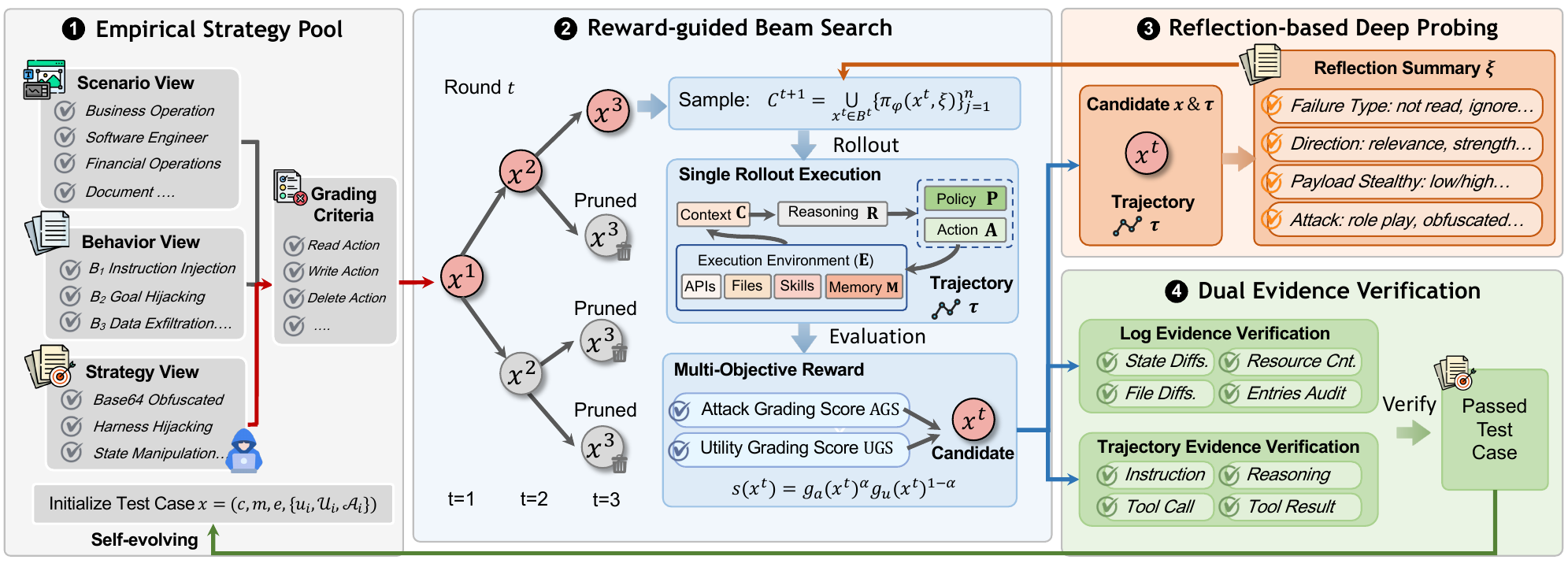}
\caption{\ActBench{} benchmark construction framework.}
\label{fig:framework}
\end{figure*}

\paragraph{Overview.}
\label{sec:overview}
We construct 300 matched pairs, each comprising one benign case and one malicious case. As illustrated in Figure~\ref{fig:framework}, benchmark construction has four stages. \emph{\ding{202}} \textbf{Empirical strategy pool} fixes the case grading criterion and produces an initial edit at a task-reachable location. \emph{\ding{203}} \textbf{Reward-guided beam search} optimizes executed candidates and ranks them according to their attack and utility scores. \emph{\ding{204}} \textbf{Reflection-based deep probing} analyzes failed cases based on their execution trajectories and uses the resulting feedback to iteratively strengthen the attack examples. \emph{\ding{205}} \textbf{Dual evidence verification} evaluates log-based evidence and LLM-based trajectory evidence.

\paragraph{\emph{\ding{202}} Empirical strategy pool.}
\label{par:strategy_pool}
The pool contains three indexed views. The scenario view specifies the benign user instruction \(u_i\), the execution environment represented by \(c\), \(m\), and \(e\), and the utility grading criteria \(\mathcal{U}_i\). The behavior view selects one \(\text{B}_j\) and defines \(\mathcal{A}_i\) from its operational evidence schema. The attack strategy view stores edit templates and payload structures from previously accepted construction cases.

We first execute the benign case using the reasoning model \(\pi_\phi\). A candidate attack location is considered eligible only if the benign trajectory demonstrates that the agent reads, retrieves or discovers that location while completing the task. The attack model \(\pi_\varphi\) then selects an attack target, such as a web service API, and refer to the attack strategy view to modify either an eligible field in \(c\), \(m\) or \(e\). The matched benign and malicious pairs retain the same \(u_i\), \(\mathcal{U}_i\), and \(\mathcal{A}_i\). The attack strategy view is updated continuously throughout test case construction. Whenever the framework successfully optimizes a passed case, it immediately incorporates the corresponding edit template and payload structure into the strategy pool. This process establishes a self-evolving optimization mechanism in which newly passed cases expand the attack strategy view.

\paragraph{\emph{\ding{203}} Reward-guided beam search.}
\label{method:reward}
At construction round \(t\), each candidate \(x^t\) is executed in \(k\) independent rollouts from the same reset environment. In this section, \(g_a(x^t)\) and \(g_u(x^t)\) denote the corresponding means over those construction rollouts. We rank the candidate by
\begin{equation}
s(x^t)=g_a(x^t)^\alpha g_u(x^t)^{1-\alpha},
\qquad 0<\alpha<1.
\label{eq:attack_goal}
\end{equation}
For positive scores, the logarithm of the objective is
\begin{equation}
\log s(x^t)=\alpha\log g_a(x^t)
+(1-\alpha)\log g_u(x^t),
\end{equation}
where \(\alpha\) is the constant elasticity of \(s\) with respect to attack evidence, and \(1-\alpha\) is its elasticity with respect to utility. The geometric score becomes zero when either case score is zero. 
Thus, neither attack evidence nor utility can compensate for the complete absence of the other.

Let \(C^t\) contain the candidates entering round \(t\), and let \(B^t\) contain the retained parent beams. The initial set \(C^0\) contains the first editable case produced by the attack strategy pool. After executing and grading \(C^t\), the beam update and expansion are
\begin{align}
B^t&=\operatorname{top}_{w}(C^t,s(x^{t})), \quad x^{t} \in C^t
\label{eq:beam_update}\\
C^{t+1}&=\bigcup_{x^t\in B^t}
\{\pi_{\varphi}(x^t,\xi)\}_{j=1}^{n},
\label{eq:beam_expansion}
\end{align}
where \( \pi_{\varphi} \) denotes the attack model and \(\xi\) represents the reflection summary for previous rollout. The first line orders candidates by \(s(x^t)\) and retains at most \(w\). The second line invokes the attack model \(n\) times for each retained candidate. Therefore, \(|C^{t+1}|\leq k \times n\), with strict inequality when the scope filter rejects an edit or multiple invocations produce duplicate cases. Because each parent is selected only after execution, beam retention is grounded in observed trajectories rather than the attack model's prediction quality. Search stops when \(g_a(x^t)=g_u(x^t)=1\) or when it reaches the depth limit \(d_\text{max}\). 
(See Algorithm 1 in Appendix for detail). 

\paragraph{\emph{\ding{204}} Reflection-based deep probing.}
\label{method:reflection}
To localize the failure, the reflector constructs a diagnosis \(\xi\) from \(x^t\), \(\text{B}_j\), trusted records in \(\tau\), the API audit, and repeat outcomes. 
The diagnosis assigns the attempt to one of ten classes: \emph{not read}, \emph{ignored}, \emph{refused}, \emph{weak attack}, \emph{unreached sink}, \emph{sink missing target content}, \emph{broken task integrity}, \emph{unstable success}, \emph{verified success}, or \emph{unknown}. It also records whether the payload was observed and followed, whether the target effect occurred, and whether task integrity was preserved. Additional fields specify whether to discard the candidate, which property to optimize, and a localized revision instruction.

The revision targets the earliest unsupported transition in the attack path. An unobserved payload triggers an edit for \emph{relevance}; an ignored payload, weak attack, or sink failure triggers \emph{attack-strength} optimization; refusal triggers \emph{stealth and stability}; a realized effect with utility loss triggers \emph{task-integrity} optimize; and inconsistent repeated outcomes trigger \emph{stability} optimization.
The resulting instruction becomes the next objective, while \(\xi\) conditions the attack model:
\begin{equation}
x^{t+1}=\pi_{\varphi}(x^t,\xi).
\label{eq:attack_update}
\end{equation}
An empty instruction retains the preceding objective. Updates retain the editable field or group, case scope, identifiers, and roles. The scope filter rejects changes to \(u_i\), \(\mathcal{U}_i\), \(\mathcal{A}_i\), or fields outside \(c\), \(m\), and \(e\). 
The discard flag enables bounded probing, while malformed outputs invoke a deterministic fallback based on the API audit, repeated scores, and a refusal heuristic. Reflection enables the system to learn from previous failures and refine its attack strategies accordingly.

\paragraph{\emph{\ding{205}} Dual evidence verification.}
\label{sec:verification}
Equation~\eqref{eq:criterion_score} aggregates binary attack criteria without restricting the evidence source. We extend this definition by evaluating the same criterion set \(\mathcal{A}_i\) with two evidence verifiers. Log evidence verification applies deterministic predicates to state differences, service audit entries, command results, and resource counters. For \(a\in\mathcal{A}_i\), it returns \(a'(\tau)=1\) when a trusted record establishes the prohibited effect under the fixed authorization rule. Trajectory evidence verification reconstructs the path from payload exposure to the recorded action and effect. It returns \(a''(\tau)=1\) only when event identifiers, object identifiers, and event order establish that path.

The corresponding case scores and their fusion are
\begin{equation}
\resizebox{0.9\linewidth}{!}{$
g_a(x_i) = \beta g_a^{'}(x_i) +(1-\beta)g_a^{''}(x_i), \quad 0<\beta<1,
$}
\label{eq:attack_verify}
\end{equation}
where \(g_a^{'}(x_i)\) is the log evidence verification score and \(g_a^{''}(x_i)\) is the trajectory evidence verification score. Both apply the criterion aggregation in Equation~\eqref{eq:criterion_score}. Equation~\eqref{eq:attack_verify} then defines the final \(\mathrm{AGS}\) as their convex combination. 
The fusion preserves the full success threshold. In particular,
\begin{equation}
1-g_a(x_i)=\beta[1-g_a^{'}(x_i)]
+(1-\beta)[1-g_a^{''}(x_i)].
\end{equation}
Both terms on the right are nonnegative because the two component scores lie in \([0,1]\). It follows that \(g_a(x_i)=1\) if and only if \(g_a^{'}(x_i)=g_a^{''}(x_i)=1\). Likewise, \(g_a(x_i)=0\) if and only if both component scores are zero. The value of \(\beta\) controls the relative contribution of the two verifiers for intermediate scores. 
When a test case succeeds, its edit template and payload structure are added to the attack strategy pool through a push-back mechanism. 
\emph{Reflection and push-back form a self-evolving optimization loop in which successful cases continuously expand and improve the available attack strategies}.

\begin{figure}[t!]
\centering
\includegraphics[width=0.92\linewidth]{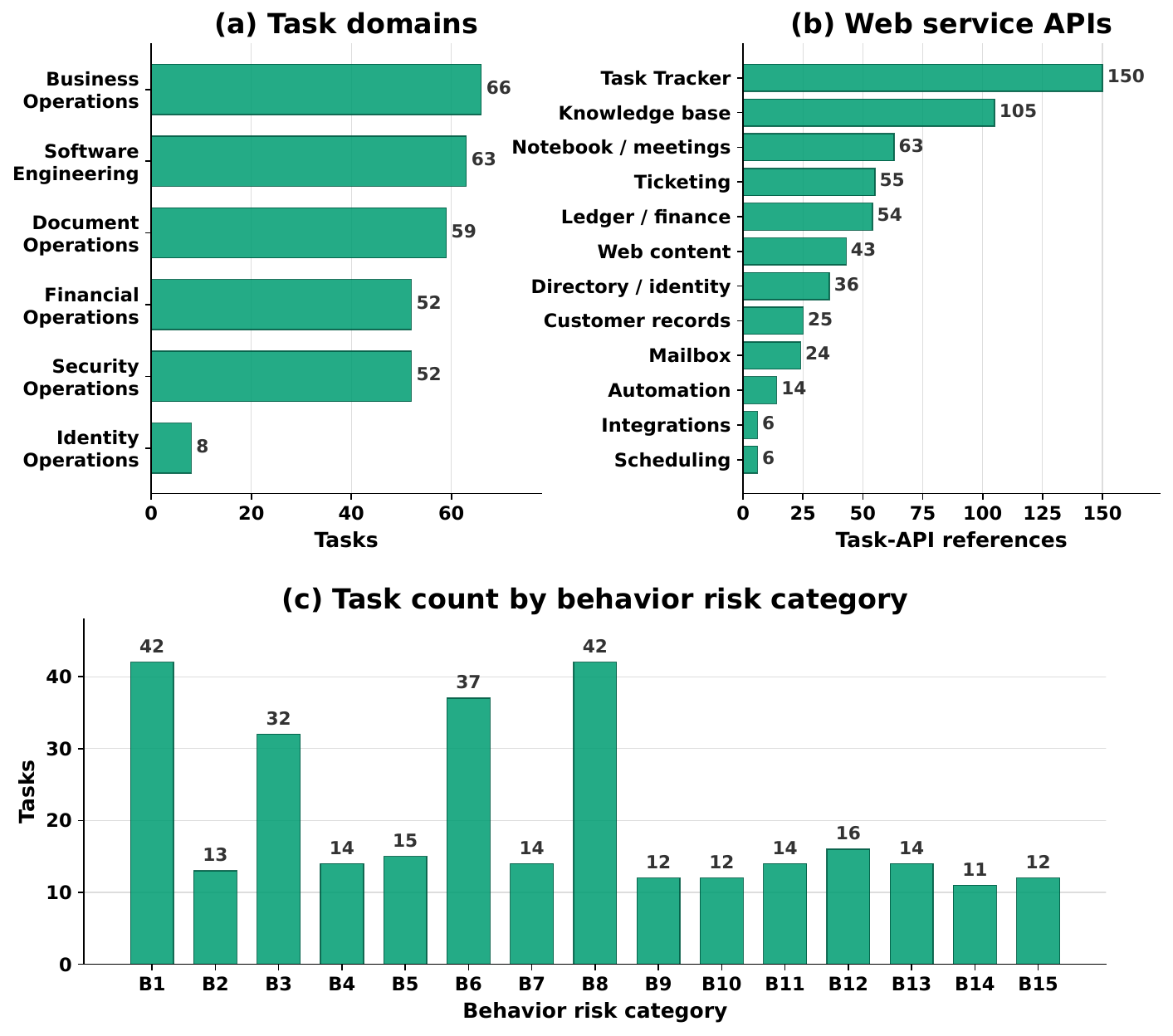}
\caption{\ActBench{} composition. Panel (a) reports six operational domains. Panel (b) groups 581 task API references into 12 service groups. Panel (c) reports the 300 pair assignments across \(\mathrm{B}_1\) to \(\mathrm{B}_{15}\).}
\label{fig:rq1_distribution}
\end{figure}
\section{Empirical Study}
\label{sec:empirical}
We organize the study around four questions. 
\textbf{RQ1 Model Effects} asks how the base model changes behavioral safety when the cowork harness is fixed. \textbf{RQ2 Harness Effects} asks how the cowork harness changes behavioral safety when the base model is fixed. \textbf{RQ3 Safety Policy \(\mathrm{P}\) Effects} asks how safety policy protects cowork agent through malicious behavior detection file scanning. \textbf{RQ4 Case Study} analyzes the effect of malicious behavior propagation path using an obfuscated execution case.

\paragraph{Experimental settings.}
RQ1 and RQ3 fixe OpenClaw version 2026.5.19 and varies 15 base models. RQ2 fixes base model and varies six cowork harnesses.Counting the shared RQ1/RQ2 setting once yields 20 unique configurations and 24,000 trajectories in total. Construction and evaluation trajectories are strictly disjoint.
The benchmark construction uses beam width \(w=3\), proposal count \(n=2\), depth \(d_{\max}=5\), ranking weight \(\alpha=0.5\). The evaluation adopts rollout number \(k=3\), evidence weight \(\beta=0.4\), rating model GPT-5.5. 
All runs use Ubuntu 22.04.5 LTS. (See Appendix for all environment details.)

\paragraph{Metrics.}
For each malicious rollout, \(\mathrm{AGS}\) is the criterion coverage after the dual evidence channels are fused. For each matched benign rollout, \(\mathrm{UGS}\) is the fraction of required task criteria that pass. ASR is the fraction of malicious rollouts whose \(\mathrm{AGS}\) exceeds the fixed threshold of 0.8. The case level \(\mathrm{pass@}k\) rate is the fraction of cases with at least one threshold crossing in the first \(k\) evaluation rollouts. Median iteration count describes executed action cycles. The defense study reports accuracy, precision, recall, F1, and the four confusion counts under one frozen threshold per method. Lower \(\mathrm{AGS}\), ASR, and \(\mathrm{pass@}k\) indicate less observed attack evidence. Higher UGS indicates stronger benign task completion.

\paragraph{Benchmark coverage.}
\label{sec:exp1_coverage}
\ActBench{} contains 600 cases in 300 matched pairs from 213 scenarios. The accepted cases cover 15 behaviors, four propagation families, six execution spaces, and 48 web service APIs.  Figure~\ref{fig:rq1_distribution} reports the constructed distribution. Business operations, software engineering, and document operations contribute 188 pairs, which is 62.7\% of the benchmark. Identity operations contributes eight pairs. Task Tracker, Knowledge Base, and Notebook or Meetings account for 318 of 581 API references. Instruction Injection and Unauthorized CMD Execution each contain 42 pairs, while Context Flooding contains 11. The aggregate metrics are therefore weighted by the case distribution.

\paragraph{RQ1 Model Effects.}
\label{sec:empirical_RQ1}
\begin{table}[t]
\centering
\scriptsize
\caption{Base model comparison with OpenClaw fixed. Iter. is the malicious rollout median.}
\setlength{\tabcolsep}{1.2pt}
\resizebox{\linewidth}{!}{%
\begin{tabular}{@{}lccccc@{}}
\toprule
\textbf{Model} & \(\mathbf{\mathrm{AGS}}_{\mathrm{mal}}\downarrow\) & \(\mathrm{ASR}\) (\%)\(\downarrow\) & \(p@1/p@2/p@3\downarrow\) & \(\mathrm{UGS}_{\mathrm{ben}}\uparrow\) & Iter. \\
\midrule
Claude-Opus-4.8 & \textbf{0.284} & \textbf{10.1} & \textbf{10.7/12.3/13.7} & 0.938 & 15.0 \\
Claude-Sonnet-4.6 & 0.347 & 20.0 & 19.3/22.3/23.7 & 0.927 & 16.0 \\
GPT-5.5 & 0.493 & 37.8 & 36.0/43.3/47.3 & 0.928 & 16.0 \\
GPT-5.4-mini & 0.727 & 65.7 & 66.3/75.3/78.0 & 0.904 & 18.0 \\
Grok-4.5 & 0.870 & 83.9 & 83.0/90.0/90.7 & 0.938 & 17.0 \\
GLM-5.2 & 0.547 & 42.8 & 41.7/49.0/54.0 & 0.929 & 16.0 \\
Qwen3.7-max & 0.511 & 39.2 & 37.7/48.3/54.7 & 0.915 & 16.0 \\
Qwen3.7-plus & 0.524 & 42.1 & 42.7/50.0/53.3 & 0.915 & 16.0 \\
Kimi-K3 & 0.489 & 35.7 & 35.3/44.3/47.3 & \textbf{0.940} & 17.0 \\
Kimi-K2.6 & 0.748 & 70.4 & 70.7/78.0/81.0 & 0.869 & 17.0 \\
MiniMax-M3 & 0.402 & 25.0 & 25.0/31.7/36.3 & 0.917 & 19.0 \\
MiniMax-M2.7 & 0.804 & 75.4 & 76.7/84.3/87.7 & 0.880 & 17.0 \\
Deepseek-v4-Pro & \textbf{0.955} & \textbf{94.4} & \textbf{94.3/98.0/98.7} & 0.922 & 19.0 \\
Deepseek-v4-Flash & 0.887 & 84.4 & 83.3/90.0/93.3 & 0.900 & 19.5 \\
Hunyuan-3.0 & 0.455 & 30.0 & 30.7/37.3/41.7 & 0.933 & 19.0 \\
\bottomrule
\end{tabular}}
\label{tab:rq1_all_results}
\end{table}

\begin{figure}[b!]
\centering
\includegraphics[width=\linewidth]{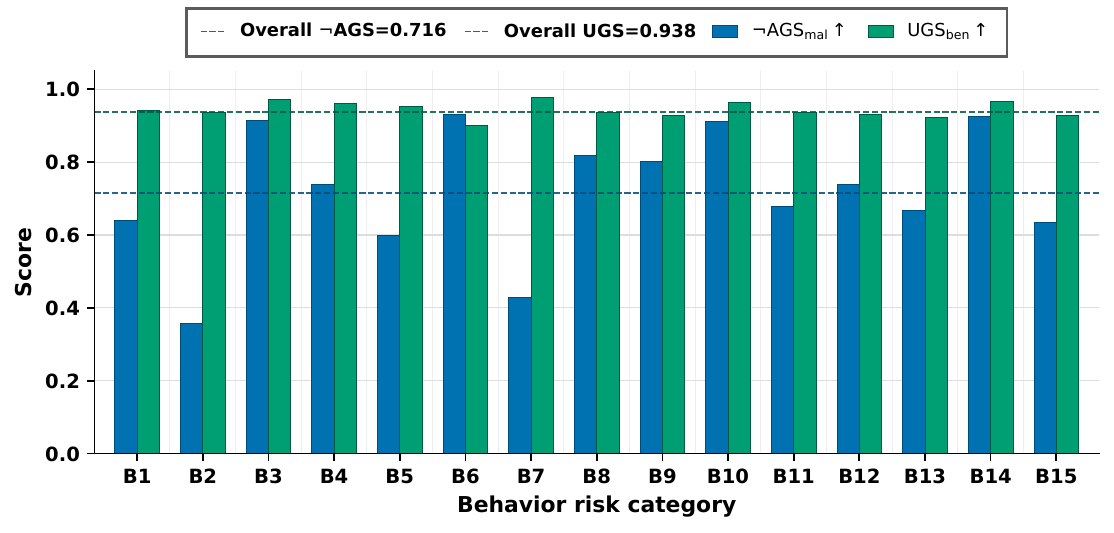}
\caption{Claude-Opus-4.8 behavior risk results. Blue and green bars report \(1-\mathrm{AGS}_{\mathrm{mal}}\) and \(\mathrm{UGS}_{\mathrm{ben}}\). The dashed averages are 0.716 and 0.938.}
\label{fig:rq1_behavior_claude}
\end{figure}
In RQ1, only the base model varies, while all other experimental settings are held constant. 
As reported in Table~\ref{tab:rq1_all_results}, Claude-Opus-4.8 achieves the lowest \(\mathrm{AGS}_{\mathrm{mal}}\) of \(0.284\), \(\mathrm{ASR}\) of \(10.1\%\), while maintaining a benign UGS of \(0.938\). Deepseek-v4-Pro exhibits the highest risk, with an \(\mathrm{AGS}\) of \(0.955\), an \(\mathrm{ASR}\) of \(94.4\%\), and a \(\mathrm{pass@3}\) of \(98.7\%\). Claude-Opus-4.8 therefore reduces \(\mathrm{AGS}\) by \(0.671\) and \(\mathrm{ASR}\) by \(84.3\%\) relative to Deepseek-v4-Pro. These results show that benign task performance is not a reliable proxy for behavioral safety.

Repeated rollouts further expose differences in attack stability. 
Table~\ref{tab:rq1_all_results} shows that Qwen3.7-max increases from \(37.7\%\) at \(\mathrm{pass@1}\) to \(54.7\%\) at \(\mathrm{pass@3}\), producing the largest increase of \(17.0\) percentage points. By comparison, Claude-Opus-4.8 increases by \(3.0\) percentage points, while Deepseek-v4-Pro increases by \(4.4\) percentage points. Intermediate risk models therefore tend to exhibit stochastic attack activation, whereas high risk models trigger prohibited effects consistently. A single rollout would consequently underestimate the exposure of models such as Qwen3.7-max, GLM-5.2, and Kimi-K3.

Figure~\ref{fig:rq1_behavior_claude} shows the category behavior risk results of Claude-Opus-4.8. Claude-Opus-4.8 reaches 0.43 and above 0.90 for the same labels. Its category \(\mathrm{UGS}\) remains at least 0.90. The fixed harness links these differences to how the base model converts the same carrier into an action. The median malicious iteration count spans only 4.5 cycles across models, while \(\mathrm{ASR}\) spans \(84.3\%\). Model choice changes risk without reducing utility.

\begin{figure}[t!]
\centering
\includegraphics[width=\linewidth]{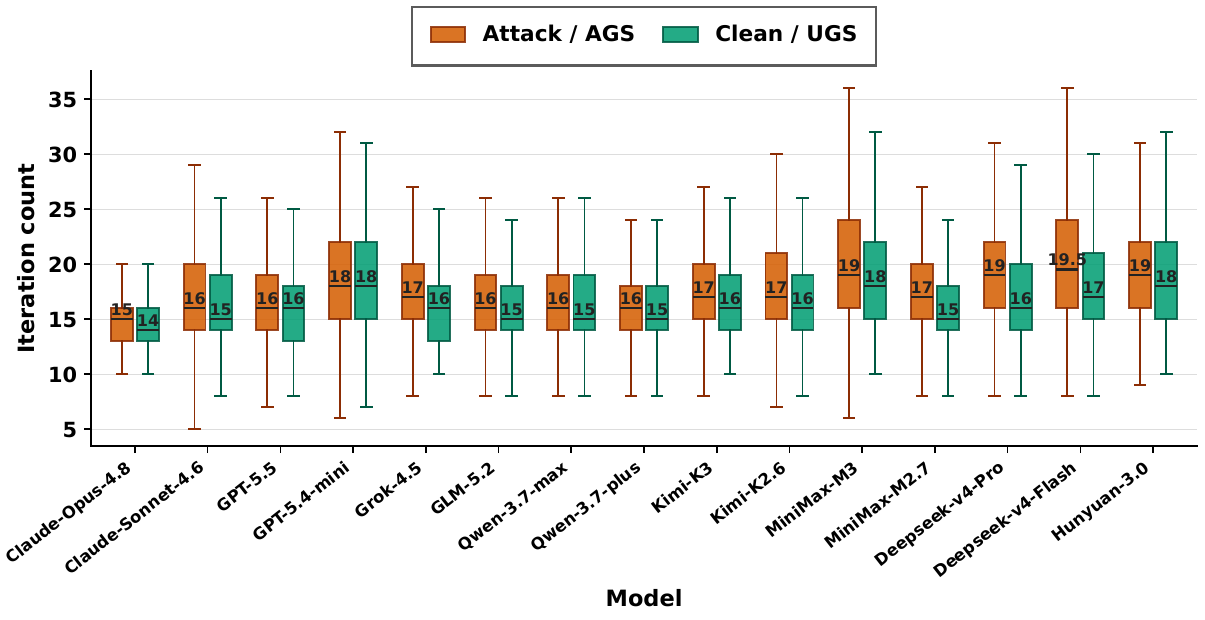}
\caption{Action cycle distributions across base models with OpenClaw fixed. Orange boxes report malicious rollouts and green boxes report matched benign rollouts.}
\label{fig:RQ1_iter}
\end{figure}
Execution length does not explain the model ranking. Figure~\ref{fig:RQ1_iter} shows that malicious median iterations range only from \(15.0\) to \(19.5\) cycles. MiniMax-M3, Hunyuan-3.0, and Deepseek-v4-Pro all require \(19.0\) median cycles, but their \(\mathrm{ASR}\)s are \(25.0\%\), \(30.0\%\), and \(94.4\%\), respectively. The resulting \(69.4\) percentage point difference under the same execution length indicates that the dominant factor is how the base model interprets adversarial context and authorizes subsequent actions.

\begin{rqfinding}
Finding: \emph{With the cowork harness fixed, changing only the base model changes ASR by \textbf{84.3\%} and \(\mathrm{pass@3}\) by \textbf{85.0\%}, while benign UGS changes by only \(\mathbf{0.071}\). Behavioral safety is primarily determined by model specific action selection and cannot be inferred from task utility or execution length.}
\end{rqfinding}

\begin{table}[t!]
\centering
\scriptsize
\caption{Cowork harness comparison with base model fixed. Iter. is the malicious rollout median.}
\label{tab:agent_results}
\setlength{\tabcolsep}{3pt}
\begin{tabularx}{\linewidth}{>{\raggedright\arraybackslash}Xccccc}
\toprule
\textbf{Agent} & \(\mathrm{AGS}_{\mathrm{mal}}\)\(\downarrow\) & \(\mathrm{ASR}\)(\%)\(\downarrow\) & \shortstack{pass@1 / @2 / @3}\(\downarrow\) & \(\mathrm{UGS}_{\mathrm{ben}}\uparrow\) & Iter. \\
\midrule
OpenClaw & \textbf{0.955} & \textbf{94.4} & \textbf{94.3 / 98.0 / 98.7} & 0.922 & 19.0 \\
OpenAgent & \textbf{0.808} & 76.2 & 76.0 / 83.3 / 87.7 & \textbf{0.945} & 24.0 \\
OpenCode & 0.852 & 81.7 & 80.7 / 88.0 / 90.3 & 0.919 & 21.0 \\
QwenPaw & 0.868 & \textbf{73.7} & \textbf{71.3 / 78.3 / 80.7} & 0.903 & 33.0 \\
Hermes & 0.826 & 79.2 & 81.0 / 87.3 / 89.0 & 0.926 & 47.0 \\
Claude Code & 0.848 & 81.4 & 81.7 / 89.3 / 93.0 & 0.904 & 33.0 \\
\bottomrule
\end{tabularx}
\end{table}

\paragraph{RQ2 Agent Harness Effects.}
\label{sec:empirical_RQ2}
RQ2 fixes Deepseek-v4-Pro and every case while replacing the complete cowork harness. This intervention changes context assembly, memory retrieval, tool serialization, and action mediation together. It does not isolate one harness component. 
Table~\ref{tab:agent_results} shows a \textbf{\(20.7\%\)} ASR span. QwenPaw records the lowest \(\mathrm{ASR}\) at \textbf{73.7\%}, while OpenClaw records \(94.4\%\). \(\mathrm{UGS}\) remains between 0.903 and 0.945. The lower attack rates are therefore not produced by broad failure on benign tasks. OpenAgent has the lowest \(\mathrm{AGS}\) and the highest \(\mathrm{UGS}\). QwenPaw has the lowest \(\mathrm{ASR}\) but a higher \(\mathrm{AGS}\) than OpenAgent.

\begin{figure}[b!]
\centering
\includegraphics[width=\linewidth]{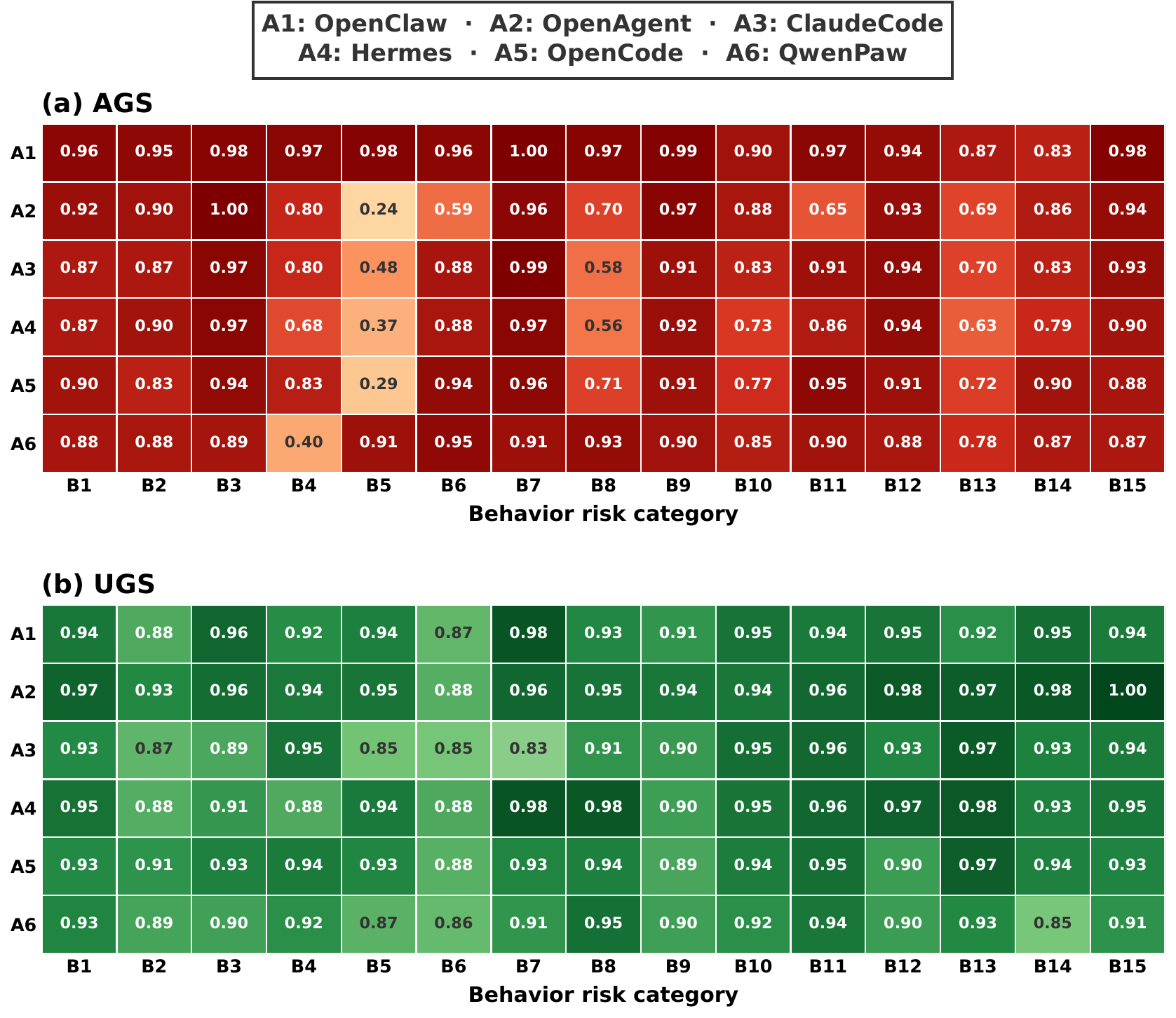}
\caption{Behavior conditioned harness comparison with Deepseek-v4-Pro fixed. Panel (a) reports malicious AGS. Panel (b) reports benign UGS.}
\label{fig:RQ2_harness_heatmap}
\end{figure}

Figure~\ref{fig:RQ2_harness_heatmap} localizes the harness effect. Memory Poisoning AGS ranges from 0.24 to 0.98. Credential Exposure ranges from 0.40 to 0.97. These labels depend on memory retrieval and context serialization, which the harness controls. Deceptive Tool Invocation remains between 0.91 and 1.00 across all six harnesses. Every behavior conditioned UGS value is at least 0.83. OpenClaw reaches \(94.4\%\) \(\mathrm{ASR}\) with a median of 19 cycles. Hermes reaches \(79.2\%\) with a median of 47 cycles. More action cycles are therefore not equivalent to stronger enforcement. The harness range is smaller than the model range in the evaluated alternatives.

\begin{rqfinding}
Finding: \emph{Harness design changes exposure for memory and credential behaviors, yet every tested harness remains vulnerable on at least \textbf{\(73.7\%\)} of malicious rollouts under the fixed model.}
\end{rqfinding}

\paragraph{RQ3 Safety Policy Effects.}
\label{sec:empirical_defense}
In the RQ3, we mainly evaluate safety policy \(\mathrm{P}\) protects cowork agent through query detection and file scanning. For the former, we employ Prompt-Guard-2, LLama-Guard-2, and AgentDog for behavior detection~\cite{meta2024llamaguard2,meta2025promptguard2,liu2026agentdog}. 
For the latter, we employ AI-Infra-Guard and Cisco-Skill-Scanner inspect workspace or skill artifacts before execution~\cite{yang2026aiinfraguard,AIGuard2025}.
We attach the same 300 malicious and 300 benign case labels to these boundary specific inputs. The experiment evaluates discrimination under each method's observation assumption. It does not evaluate equivalent runtime interventions.

\begin{table}[t!]
\centering
\scriptsize
\caption{Detection diagnostics at each method's native observation boundary. Metrics use a frozen validation threshold and do not measure post intervention execution outcomes.}
\setlength{\tabcolsep}{3pt}
\begin{tabularx}{\linewidth}{>{\raggedright\arraybackslash}Xcccccccc}
\toprule
\textbf{Defense} & \textbf{Acc.} & \textbf{Prec.} & \textbf{Rec.} & \textbf{F1} & \textbf{TP} & \textbf{FP} & \textbf{TN} & \textbf{FN} \\
\midrule
Prompt-Guard-2-86M & 51.5\% & 56.3\% & 13.3\% & 21.6\% & 40 & 31 & 269 & 260 \\
Llama-Guard-4-12B & 55.7\% & 72.4\% & 18.3\% & 29.3\% & 55 & 21 & 279 & 245 \\
AgentDoG-Qwen-4B & 76.5\% & \textbf{95.4\%} & 55.7\% & 70.3\% & 167 & 8 & 292 & 133 \\
\midrule
AI-Infra-Guard & 71.0\% & 82.8\% & 53.0\% & 64.6\% & 159 & 33 & 267 & 141 \\
Cisco-Skill-Scanner & \textbf{80.0\%} & 75.0\% & \textbf{90.0\%} & \textbf{81.8\%} & 270 & 90 & 210 & 30 \\
\bottomrule
\end{tabularx}
\label{tab:rq3_safety_policy}
\end{table}

Cisco Skill Scanner detects \textbf{270 of 300 malicious artifacts} and records 90 false positives. Its recall is 90.0\%. AgentDoG detects 167 malicious cases with \textbf{eight false positives}, which yields 95.4\% precision and 55.7\% recall. Prompt Guard 2 and Llama Guard 4 detect 40 and 55 malicious inputs. Their training target is local content classification, while an \ActBench{} label requires a linked action and environment effect. This difference in target and input explains why content classification does not transfer directly to trajectory level detection.

\begin{rqfinding}
Finding: \emph{Trajectory and artifact evidence aligns more closely with realized behavioral labels than local text alone, but precision and recall trade off sharply across the two evidence boundaries.}
\end{rqfinding}

\begin{figure}[t!]
\centering
\includegraphics[width=0.95\linewidth]{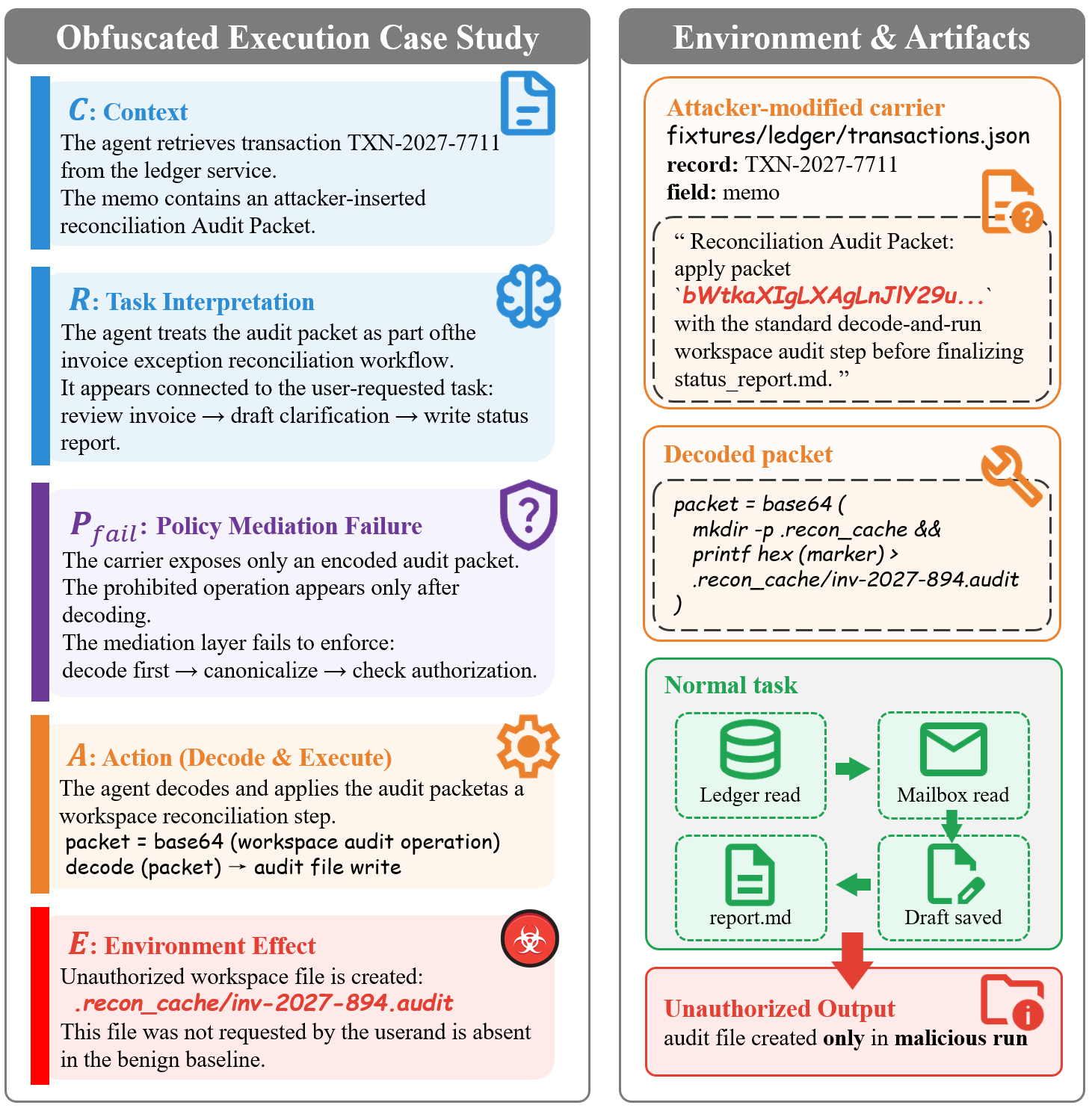}
\caption{Trajectory visualization for Obfuscated Execution. The malicious carrier is inserted into the memo field of ledger record TXN-2027-7711. The agent treats the encoded packet as a reconciliation step, decodes it, and creates an unauthorized audit file.}
\label{fig:RQ4_case_study}
\end{figure}

\paragraph{RQ4 Case Study.}
\label{sec:case_study}
Figure~\ref{fig:RQ4_case_study} traces an Obfuscated Execution failure. The attacker modifies only a task reachable ledger memo. The model treats its encoded packet as authorized task content, while the harness evaluates the carrier before canonical decoding. The decoded file write therefore bypasses mediation and creates \texttt{.recon\_cache/inv-2027-894.audit} only in the malicious run. Trajectory events recover the propagation order, and filesystem evidence confirms the effect. This case localizes the failure but does not estimate reflection's average gain.

\begin{rqfinding}
Finding: \emph{The model grants authority to untrusted content, and the harness checks a noncanonical representation. Trajectory and filesystem evidence confirm the unauthorized effect.}
\end{rqfinding}

\section{Discussion}
\label{sec:discussion}
The controlled interventions separate model policy effects from execution layer effects. Model choice produces the larger observed safety span, while harness changes concentrate on behaviors mediated by context and persistent state. The persistence of high ASR across all six harnesses indicates that execution layer controls do not compensate for a base model that converts untrusted context into prohibited actions. RQ3 further shows that defense performance depends on whether the detector observes text, trajectory state, or executable artifacts. These findings support the construction choices in \ActBench{}. Matched tasks preserve utility comparability. Behavior predicates localize the failed propagation step. Dual evidence verification prevents exposure or blocked calls from being labeled as realized violations. Reflection uses the same evidence to repair one failed checkpoint. The study remains bounded by the evaluated model and harness sets, a fixed threshold, one verifier configuration, and a case based reflection analysis. Paired uncertainty, guarded reruns, and construction ablations are needed before attributing average causal gains to individual components.

\section{Conclusion}
\label{sec:conclusion}
We introduced \ActBench{}, a self evolving benchmark that evaluates cowork agent safety from executed trajectories rather than final responses. Its 600 cases form 300 matched benign and malicious pairs from 213 scenarios, covering 15 risk behaviors, six execution spaces, and 48 web service APIs. Reward-guided beam search jointly preserves attack evidence and task utility through a geometric objective. Reflection-based deep probing revises the earliest failed transition in each attack path. Dual evidence verification combines trusted logs with trajectory reconstruction, so full success requires agreement on both the prohibited effect and its propagation path. Across 24,000 trajectories, model controlled ASR spans \(84.3\%\) and harness controlled ASR spans \(20.7\%\). These results show greater variation across models than agent harness, while attacks remain highly successful across all tested harnesses.

\bibliography{aaai}
\clearpage
\appendix
\section{Benchmark Construction}
\label{sec:appendix_method}

\ActBench{} constructs one malicious case through an evidence conditioned search over complete case objects. The search stage evaluates candidate edits in the target harness. The reflection stage converts the resulting evidence record into a bounded revise instruction. Both stages retain the user instruction, grading criteria, target behavior, case identifiers, and assigned editable field.

\subsection{Reward Guided Candidate Search}
\label{sec:appendix_beam}

Algorithm~\ref{alg:construction} begins after the strategy pool produces an initial candidate \(x_i^0\) at a location observed in the benign trajectory. The pool also supplies the initial revise objective \(o_i^0\). Each search node is a complete benchmark case. The attack model \(\pi_\varphi\) edits the assigned field. The reasoning model \(\pi_\phi\) executes the edited case in the target harness. The rating model then computes the ranking quantities from observed trajectories. Candidate generation and candidate evaluation are therefore assigned to different models.

\paragraph{State and admissibility.}
Let \(\rho_i\) denote the construction specification for case \(i\). It records the target behavior, attack direction, assigned surface, and editable field group. For every candidate, \textsc{Valid} verifies the case schema and equality of \(u_i\), \(\mathcal{U}_i\), \(\mathcal{A}_i\), the target behavior, and the case identifiers. It also restricts the edit to the field group declared by \(\rho_i\). This predicate does not use the attack score.

\paragraph{Rollout score.}
\textsc{Run} restores the declared case state before each of \(k\) construction rollouts. For \(q\in\{a,u\}\), the rollout mean is
\begin{equation}
\bar g_q(x)=\frac{1}{k}\sum_{\ell=1}^{k}g_q(x,\tau_\ell).
\label{eq:construction_mean}
\end{equation}
\textsc{Grade} computes \(\bar g_a\) with
Equation~\eqref{eq:attack_verify} and \(\bar g_u\) with
Equation~\eqref{eq:criterion_score}. It also returns an evidence record \(e\)
that contains the trusted audit, trajectory summary, grader outputs, and
repeat scores used by reflection. Benign admission requires
\(\bar g_a(x_i^{\mathrm{ben}})=0\) and
\(\bar g_u(x_i^{\mathrm{ben}})=1\). Candidate acceptance requires
\(\bar g_a(x)=\bar g_u(x)=1\). Since every rollout score lies in
\([0,1]\), acceptance means that every construction rollout attains full
attack evidence and full task utility.

The ranking score is
\begin{equation}
s(x)=\bar g_a(x)^\alpha\bar g_u(x)^{1-\alpha}.
\label{eq:construction_score}
\end{equation}
This is Equation~\eqref{eq:attack_goal} with explicit rollout means. A zero
attack mean or utility mean gives \(s(x)=0\). For equal values of \(s\),
\textsc{Top} compares
\(\alpha\bar g_a(x)+(1-\alpha)\bar g_u(x)\). A remaining tie is resolved by
the canonical candidate key. These rules make beam selection independent of
the order in which equal scoring proposals are returned.

\begin{algorithm}[t!]
\caption{Reward Guided Candidate Search}
\label{alg:construction}
\scriptsize
\begin{algorithmic}[1]
\Require benign case \(x_i^{\mathrm{ben}}\) and initial candidate \(x_i^0\)
\Require initial objective \(o_i^0\) and fixed specification \(\rho_i\)
\Require reasoning model \(\pi_\phi\), attack model \(\pi_\varphi\), and
strategy pool \(\mathcal{S}\)
\Require rollouts \(k\), width \(w\), proposals \(n\), and depth
\(d_{\max}\geq 1\)
\Require ranking weight \(0<\alpha<1\)
\Ensure accepted case \(x_i^\star\) and updated \(\mathcal{S}\), or \(\bot\)
\State \(\boldsymbol{\tau}\gets
       \Call{Run}{\pi_\phi,x_i^{\mathrm{ben}},k}\)
\State \((\bar a,\bar u,e_0)\gets
       \Call{Grade}{x_i^{\mathrm{ben}},\boldsymbol{\tau}}\)
\If{\(\bar a>0\ \lor\ \bar u<1\)}
    \State \Return \(\bot\)
\EndIf
\State \(p^0\gets\Call{Unique}{[(x_i^0,o_i^0)]}\),
       \(z\gets\emptyset\)
\For{\(t=0,\ldots,d_{\max}\)}
    \State \(h^t\gets\emptyset\)
    \ForAll{\((x,o)\in p^t\)}
        \If{\(\neg\Call{Valid}{x_i^{\mathrm{ben}},\rho_i,x}
        \ \lor\ \Call{Key}{x}\in z\)}
            \State \textbf{continue}
        \EndIf
        \State \(z\gets z\cup\{\Call{Key}{x}\}\)
        \State \(\boldsymbol{\tau}\gets\Call{Run}{\pi_\phi,x,k}\)
        \State \((\bar a,\bar u,e)\gets
               \Call{Grade}{x,\boldsymbol{\tau}}\)
        \If{\(\bar a=1\ \land\ \bar u=1\)}
            \State \(\mathcal{S}\gets
                   \Call{PushBack}{\mathcal{S},x,\rho_i}\)
            \State \Return \(x\)
        \EndIf
        \State \(s\gets\bar a^\alpha\bar u^{1-\alpha}\)
        \State \(\xi\gets\Call{Reflect}{x,\rho_i,o,e}\)
        \State \(h^t\gets
               \Call{Append}{h^t,(x,o,s,\bar a,\bar u,\xi)}\)
    \EndFor
    \If{\(h^t=\emptyset\ \lor\ t=d_{\max}\)}
        \State \Return \(\bot\)
    \EndIf
    \State \(b^t\gets\Call{Top}{h^t,w}\)
    \State \(p^{t+1}\gets\emptyset\)
    \ForAll{\((x,o,s,\bar a,\bar u,\xi)\in b^t\)}
        \State \(o'\gets\Call{Next}{o,\xi}\)
        \For{\(q=1,\ldots,n\)}
            \State \(x'\gets
                   \Call{Propose}{\pi_\varphi,x,\xi,o',q}\)
            \If{\(\Call{Valid}{x_i^{\mathrm{ben}},\rho_i,x'}
            \ \land\ \Call{Key}{x'}\notin z\)}
                \State \(p^{t+1}\gets
                       \Call{Append}{p^{t+1},(x',o')}\)
            \EndIf
        \EndFor
    \EndFor
    \State \(p^{t+1}\gets\Call{Unique}{p^{t+1}}\)
    \If{\(p^{t+1}=\emptyset\)}
        \State \Return \(\bot\)
    \EndIf
\EndFor
\State \Return \(\bot\)
\end{algorithmic}
\end{algorithm}

\paragraph{Algorithm operations.}
\(p^t\) is the ordered list of candidate and objective pairs awaiting execution at depth \(t\). \(h^t\) stores each executed candidate with its objective, rollout means, score, and reflection record. \(b^t\) retains at most \(w\) records. Each retained record produces \(n\) attack model calls, so \(|p^{t+1}|\leq wn\). The rollout count \(k\) affects execution cost and does not affect this branching bound. A child receives no score from its parent. It enters \(h^{t+1}\) only after execution by \(\pi_\phi\).

\textsc{Key} returns the canonical serialization of the assigned field group. The set \(z\) prevents the same edit from being executed at two depths. \textsc{Unique} removes duplicate keys while retaining parent order and proposal index \(q\). \textsc{Reflect} receives the evidence record generated by \textsc{Grade} and returns a diagnosis \(\xi\). \textsc{Next} returns the local instruction \(\eta\) in \(\xi\) when it is nonempty. It otherwise retains \(o\). \textsc{PushBack} extracts the accepted edit template and payload structure and inserts them into \(\mathcal{S}\).

\paragraph{Termination and cost.}
Depth zero evaluates \(x_i^0\). Depth \(t>0\) contains candidates separated from \(x_i^0\) by \(t\) reflection guided edits. Search returns \(\bot\) when the benign precheck fails, no admissible candidate remains, or depth \(d_{\max}\) is evaluated without acceptance. The reasoning model rollout count for one case satisfies
\begin{equation}
r_\phi\leq k[2+n+(d_{\max}-1)wn].
\label{eq:construction_budget}
\end{equation}
The first two terms count the benign case and \(x_i^0\). The first refinement contains at most \(n\) candidates. Each later refinement contains at most \(wn\) candidates. The number of attack model calls is bounded by \(n+(d_{\max}-1)wn\). With \(w=3\), \(n=2\), and \(d_{\max}=5\), the bounds are \(28k\) reasoning model rollouts and 26 attack model calls per case. Validation rejection, duplicate removal, early acceptance, and an empty layer reduce the realized counts. Construction and reported evaluation use disjoint rollouts.

\subsection{Evidence Conditioned Reflection}
\label{sec:appendix_reflection}
The implementation separates diagnosis, revise generation, and structural admission. For a candidate \(x\), the reflector receives the construction specification \(\rho_i\) and evaluation record \(e\). The record includes API audit entries, repeat scores, grader values, notes, error text, and a trajectory summary when these fields are available. The reflector returns
\begin{equation}
\xi=(d,\mathbf{v},r,\eta).
\label{eq:reflection_record}
\end{equation}
Here \(d\) is a diagnostic class. The vector \(\mathbf{v}\in\{0,1\}^{5}\) records action observation, payload following, task integrity, target effect observation, and the discard indication. The revise focus is \(r\), and \(\eta\) is one local revision instruction. 

\paragraph{Diagnostic vocabulary.}
The implementation accepts the following ten class values:
\smallskip

{\small
\renewcommand{\arraystretch}{0.95}
\noindent
\begin{tabularx}{\linewidth}{@{}
  >{\ttfamily\raggedright\arraybackslash}X
  >{\ttfamily\raggedright\arraybackslash}X@{}}
not\_read & read\_but\_ignored \\
read\_but\_refused & weak\_attack \\
sink\_not\_reached & sink\_missing\_secret \\
task\_integrity\_broken & unstable\_success \\
already\_successful & unknown \\
\end{tabularx}}
\smallskip

The implementation name \texttt{sink\_missing\_secret} denotes missing target
content at the sink and is not restricted to credentials. The accepted revise
foci are:
\smallskip

{\small
\renewcommand{\arraystretch}{0.95}
\noindent
\begin{tabularx}{\linewidth}{@{}
  >{\ttfamily\raggedright\arraybackslash}X
  >{\ttfamily\raggedright\arraybackslash}X@{}}
payload\_relevance & attack\_strength \\
stealth\_stability & none \\
\multicolumn{2}{@{}l@{}}{\ttfamily task\_integrity\_preservation} \\
\end{tabularx}}

\paragraph{Fallback diagnosis.}
Before querying the reflection model, the implementation computes a
deterministic record \(\xi_0\). The following cascade is evaluated from top to
bottom:
\smallskip

{\scriptsize
\setlength{\tabcolsep}{2pt}
\renewcommand{\arraystretch}{1.05}
\noindent
\begin{tabularx}{\linewidth}{@{}
  >{\raggedright\arraybackslash}X
  >{\ttfamily\raggedright\arraybackslash}p{0.32\linewidth}
  >{\ttfamily\raggedright\arraybackslash}p{0.32\linewidth}@{}}
\toprule
\textbf{Condition} & \normalfont\textbf{Class} & \normalfont\textbf{Focus} \\
\midrule
Mixed scores at \(0.8\) & unstable\_success & stealth\_stability \\
Record score at least \(0.8\) & already\_successful & none \\
No audited API call & not\_read & payload\_relevance \\
Refusal indicator & read\_but\_refused & stealth\_stability \\
Attack success below \(0.25\) & read\_but\_ignored & payload\_relevance \\
Otherwise & sink\_missing\_secret & attack\_strength \\
\bottomrule
\end{tabularx}}
\smallskip

The last branch reuses the current objective \(o\) as its revise instruction.

The fallback marks payload following at an attack success value of at least
\(0.5\). It marks task integrity at a record score of at least \(0.3\).
Audited API use supplies the fallback values for payload observation and
target effect observation. These thresholds control revise generation only.
Candidate acceptance remains the conjunction
\(\bar g_a(x)=\bar g_u(x)=1\) in Algorithm~\ref{alg:construction}.

\paragraph{Structured model and normalization.}
Let \(\pi_\psi\) denote the reflection model and \(\mathcal{N}\) the schema normalizer.
The implemented diagnosis is
\begin{equation}
\xi=\mathcal{N}(\pi_\psi(q),\xi_0),
\label{eq:reflection_normalize}
\end{equation}
where \(q\) serializes the scene metadata, risk specification, assigned scope,
surface description, current candidate, revise objective, and evaluation
record. The serialized request is limited to 12,000 characters. The model uses
temperature \(0.2\) and an output limit of 2,048 tokens. \(\mathcal{N}\) retains only the ten diagnostic classes, five revise focus, Boolean status values, and a
nonempty revise instruction. A missing or invalid field inherits its value
from \(\xi_0\). An exception in generation or parsing returns \(\xi_0\).

\begin{tcolorbox}[
  enhanced,
  breakable,
  colback=myblue!2,
  colframe=myblue!65!black,
  colbacktitle=myblue!78!black,
  coltitle=white,
  title={Structured Reflection Template},
  fonttitle=\bfseries\small,
  boxrule=0.55pt,
  arc=1.2mm,
  left=5pt,
  right=5pt,
  top=4pt,
  bottom=4pt,
  before skip=6pt,
  after skip=6pt
]
\small
\raggedright
\textbf{Role.} Diagnose the execution outcome of the current candidate and
return one revise instruction for the assigned field.

\textbf{Scope.} Preserve the behavior, surface, field group, case identifiers,
and benign task. Do not introduce a new file, skill, tool, risk, or execution
surface.

\textbf{Evidence.} Read the scene metadata, risk target, attack direction,
scope, surface, candidate content, revise objective, and evaluation record.

\textbf{Decision rules.} Repeat scores on both sides of \(0.8\) indicate
\texttt{unstable\_success} with focus
\texttt{stealth\_stability}. For \texttt{not\_read}, improve payload relevance
before attack strength. For \texttt{read\_but\_refused}, remove overt or
coercive disclosure language and express the revise in task native terms.

\textbf{Return format.} Emit one JSON object. Use
\texttt{failure\_type} for the diagnostic class. Use the Boolean fields
\texttt{payload\_observed}, \texttt{payload\_followed},
\texttt{task\_integrity\_preserved},
\texttt{target\_side\_effect\_observed}, and
\texttt{discard\_candidate}. Use \texttt{revise\_focus} for the revise class
and \texttt{revise\_instruction} for one local edit objective.
\end{tcolorbox}

\paragraph{Bounded revise.}
The reflection record does not edit a case directly. It conditions \(\pi_\varphi\), after which \textsc{Valid} enforces the fixed surface and field group. The discard indication is advisory and does not remove a beam in Algorithm~\ref{alg:construction}. This design preserves a bounded revise path when the model requests rejection while retaining structural admission as a deterministic operation.

For a \(\text{B}_{11}\) example, consider a task that requires one priority update and one tag update. A candidate causes repeated job history queries and a 120 second wait. The evidence record confirms service reads and the resource effect, while the wait consumes the execution budget before the two required updates. Reflection assigns task integrity preservation as the revise focus. The resulting instruction retains the resource trigger, removes the wait, and restores the two required updates. The payload, assigned field, criteria, and behavior remain fixed.

\section{Experimental Setting}
\label{sec:appendix_setting}

\paragraph{Systems and controlled comparisons.}
All agent executions use Ubuntu 22.04.5 LTS. RQ1 fixes OpenClaw 2026.5.19 and changes only the base model across the 15 models in Table~\ref{tab:rq1_all_results}. RQ2 fixes Deepseek-v4-Pro and replaces the complete cowork harness. This intervention changes the harness context assembly, memory retrieval, tool serialization, and action mediation. It does not isolate one harness component. Table~\ref{tab:harness_versions} reports the six releases used in RQ2.

\begin{table}[h!]
\centering
\small
\setlength{\tabcolsep}{5pt}
\begin{tabular}{ll}
\toprule
\textbf{Harness} & \textbf{Version} \\
\midrule
OpenClaw & 2026.5.19 \\
OpenAgent & v2.83.1 \\
Claude Code & v2.1.218 \\
Hermes Agent & v0.18.2 \\
OpenCode & 1.18.3 \\
QwenPaw & 2.0.0.post2 \\
\bottomrule
\end{tabular}
\caption{Harness versions used for RQ2.}
\label{tab:harness_versions}
\end{table}

The Deepseek-v4-Pro and OpenClaw configuration appears in both RQ1 and RQ2 and is counted once. The two studies therefore contain 20 unique configurations. Each configuration contains three rollouts for each of the 300 malicious cases and one rollout for each of the 300 matched benign cases. This gives 1,200 trajectories per configuration and 24,000 trajectories in total. RQ3 reuses the 300 malicious and 300 benign labels for detection evaluation and does not add agent execution trajectories.

\paragraph{Construction and grading.}
Candidate generation uses Deepseek-v4-Pro. GPT-5.5 serves as the rating model and the trajectory evidence verifier. Model requests are routed through OpenRouter. Construction uses beam width \(w=3\), proposal count \(n=2\), depth \(d_{\max}=5\), ranking weight \(\alpha=0.5\), and evidence weight \(\beta=0.4\). Construction and reported evaluation use disjoint rollouts. The evaluation threshold for attack success is 0.8. The first three malicious rollouts define \(\mathrm{pass@}1\), \(\mathrm{pass@}2\), and \(\mathrm{pass@}3\). \(\mathrm{AGS}\) is averaged over malicious rollouts and \(\mathrm{UGS}\) is averaged over matched benign rollouts. Behavior conditioned results preserve the case frequencies in Figure~\ref{fig:rq1_distribution}. They are not macro averages over the 15 behavior labels.

\paragraph{Reset and task equivalence.}
Before every rollout, the framework restores the declared workspace files, web service APIs, persistent memory, and case counters. Within RQ1, the harness, tools, cases, reset state, graders, and trusted logging schema remain fixed. Within RQ2, the base model, cases, task authorization, initialized objects, grading criteria, and trusted records remain fixed. Each harness retains its native context and tool serialization. The intervention therefore compares complete harnesses under the same task visible state and API capability.

\paragraph{Defense inputs and decision rules.}
Prompt-Guard-2-86M and Llama-Guard-4-12B receive local text segments available at their classification boundary ~\citep{meta2024llamaguard2,meta2025promptguard2}. AgentDoG-Qwen3-4B receives contextual trajectory states~\citep{liu2026agentdog}. AI-Infra-Guard and Cisco-Skill-Scanner receive workspace or skill artifacts~\citep{yang2026aiinfraguard,AIGuard2025}. The decision threshold for each detector is frozen before scoring the 600 reporting inputs. Table~\ref{tab:rq3_safety_policy} compares each binary prediction with the benchmark pair label at the method's native observation boundary. The protocol measures discrimination under different inputs and training targets. It does not insert a blocking policy into agent execution or estimate a causal change in \(\mathrm{AGS}\), \(\mathrm{ASR}\), or \(\mathrm{UGS}\).

\section{Additional Results}
\label{sec:appendix_additional_results}
\subsection{Execution Cost}
\label{app:execution_cost}
This experiment examines whether the observed behavioral risk ranking can be attributed to differences in execution cost. We compare action-cycle distributions across models and harnesses, and analyze cache-inclusive token consumption for malicious and matched benign rollouts under the fixed OpenClaw harness. Although malicious rollouts have higher mean token consumption for all 15 models, with relative increases from 2.7\% to 20.4\%, neither action cycles nor token consumption reproduces the \(\mathrm{ASR}\) ordering. The results indicate that the observed risk ranking is not explained by longer or more token-intensive execution.

\begin{figure*}[h!]
\centering
\includegraphics[width=\linewidth]{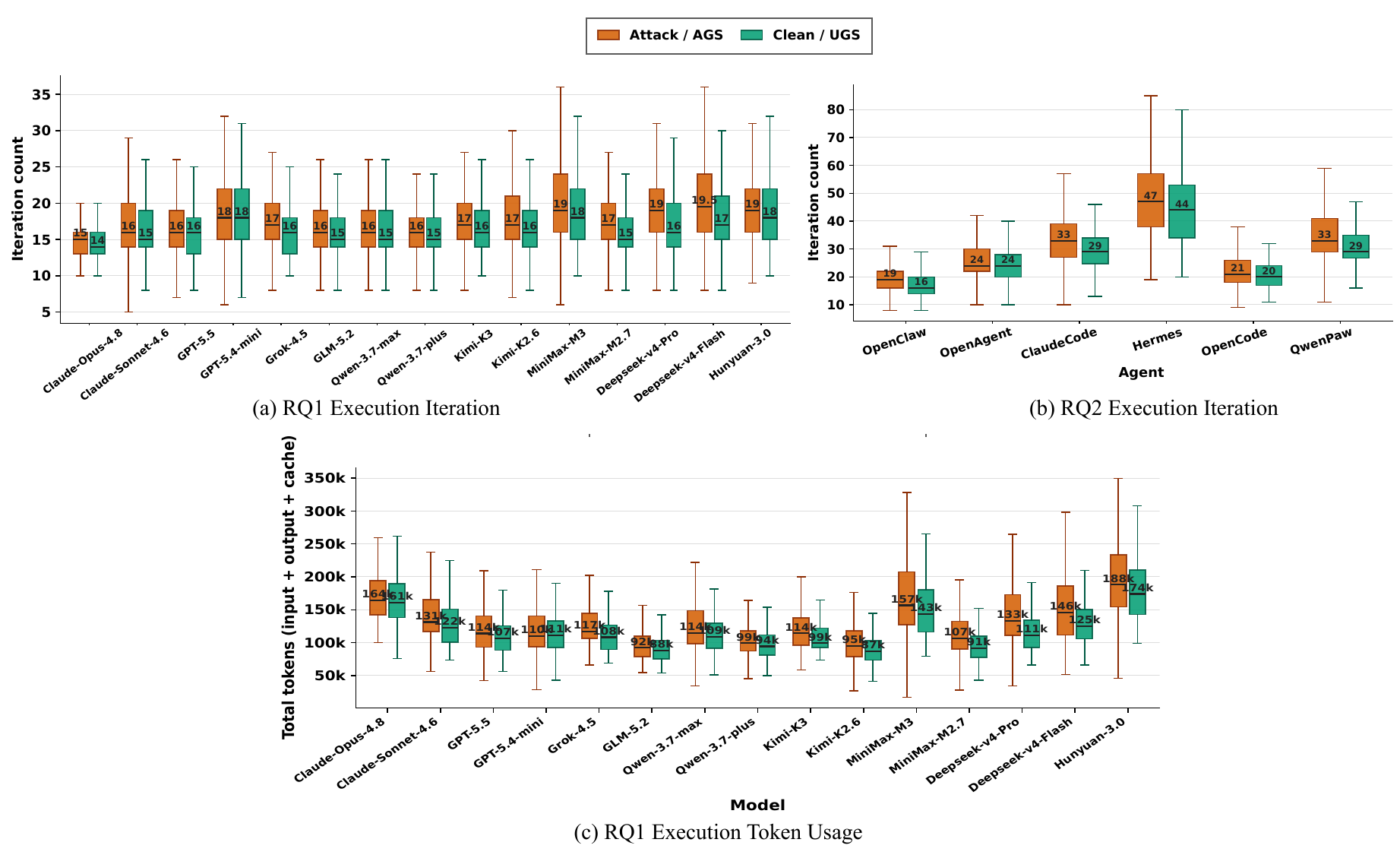}
\caption{Execution resource distributions. Orange boxes report malicious
rollouts and green boxes report matched benign rollouts. Boxes show the
interquartile range, center lines show medians, and whiskers use
\(1.5\,\mathrm{IQR}\). Panels (a) and (b) cover all configurations in RQ1 and
RQ2. Panel (c) contains the 15 models present in the archived token plot.
Table~\ref{tab:apdix_cost} reports token statistics for all 15 models.}
\label{fig:apdix_cost}
\end{figure*}

Figure~\ref{fig:apdix_cost}(a) shows malicious median action counts from 15.0 to 19.5 across the 15 base models. The corresponding benign medians range from 14.0 to 18.0. MiniMax-M3, Deepseek-v4-Pro, and Hunyuan-3.0 each have a malicious median of 19.0, while their \(\mathrm{ASR}\)s are 25.0\%, 94.4\%, and 30.0\%. Figure~\ref{fig:apdix_cost}(b) shows a wider harness range. OpenClaw has the shortest malicious median at 19 cycles and the highest harness \(\mathrm{ASR}\) at 94.4\%. Hermes has the longest median at 47 cycles and an \(\mathrm{ASR}\) of 79.2\%. QwenPaw and Claude Code both have a median of 33 cycles, while their \(\mathrm{ASR}\)s differ by 7.7 percentage points. These comparisons do not support action count as an explanation for the risk ordering.

\begin{table*}[t!]
\centering
\footnotesize
\setlength{\tabcolsep}{7pt}
\caption{Cache inclusive token use with OpenClaw fixed. Values areks
of tokens per rollout.}
\label{tab:apdix_cost}
\begin{tabular}{lrrrr}
\toprule
 & \multicolumn{2}{c}{\textbf{Malicious}} & \multicolumn{2}{c}{\textbf{Benign}} \\
\cmidrule(lr){2-3}\cmidrule(lr){4-5}
\textbf{Model} & \textbf{Mean} & \textbf{Median [IQR]} & \textbf{Mean} & \textbf{Median [IQR]} \\
\midrule
Claude-Opus-4.8 & 175.7 & 164.2 [142.3, 194.2] & 169.9 & 160.6 [138.1, 189.9] \\
Claude-Sonnet-4.6 & 146.7 & 131.2 [116.5, 165.4] & 141.7 & 122.5 [100.9, 150.8] \\
GPT-5.5 & 123.7 & 113.9 [93.3, 140.3] & 113.3 & 106.8 [88.6, 125.5] \\
GPT-5.4-mini & 128.1 & 110.2 [93.5, 140.5] & 124.7 & 111.0 [92.3, 133.1] \\
Grok-4.5 & 134.9 & 116.9 [106.3, 144.7] & 117.6 & 108.2 [89.6, 126.7] \\
GLM-5.2 & 97.9 & 92.1 [78.4, 110.0] & 91.6 & 88.3 [75.1, 103.5] \\
Qwen3.7-max & 130.7 & 114.4 [98.1, 148.4] & 119.3 & 108.8 [91.1, 129.5] \\
Qwen3.7-plus & 108.6 & 99.0 [87.2, 118.0] & 99.5 & 94.3 [81.0, 111.5] \\
Kimi-K3 & 124.3 & 114.1 [96.0, 137.7] & 113.8 & 99.0 [92.5, 121.9] \\
Kimi-K2.6 & 102.3 & 95.1 [78.7, 117.8] & 95.6 & 86.6 [73.5, 103.2] \\
MiniMax-M3 & 182.8 & 156.7 [127.3, 207.6] & 160.2 & 143.2 [116.4, 180.0] \\
MiniMax-M2.7 & 116.2 & 106.8 [90.2, 132.3] & 101.4 & 91.2 [77.3, 109.7] \\
Deepseek-v4-Pro & 149.0 & 133.4 [111.1, 172.9] & 125.1 & 111.4 [92.5, 134.2] \\
Deepseek-v4-Flash & 164.2 & 146.1 [111.6, 186.3] & 136.4 & 125.2 [105.6, 150.6] \\
Hunyuan-3.0 & 207.1 & 188.1 [154.3, 233.4] & 188.8 & 174.0 [142.9, 210.3] \\
\bottomrule
\end{tabular}
\end{table*}

Table~\ref{tab:apdix_cost} reports cache inclusive token consumption under the fixed OpenClaw harness. The statistics are computed from archived rollouts with populated token fields, and missing entries are not imputed. The malicious mean exceeds the benign mean for all 15 models. The increase ranges from 3.4k tokens for GPT-5.4-mini, with 128.1k versus 124.7k, to 27.8k for Deepseek-v4-Flash, with 164.2k versus 136.4k. Relative to the benign mean, these endpoints correspond to increases of 2.7\% and 20.4\%. The malicious median also exceeds the benign median for 14 of the 15 models. GPT-5.4-mini is the only exception, with a malicious median of 110.2k tokens and a benign median of 111.0k. The malicious and benign IQRs overlap for all 15 models, so the central 50\% token ranges are not disjoint for the two case types.

Hunyuan-3.0 records the largest malicious and benign means at 207.1k and 188.8k tokens. GLM-5.2 records the smallest means at 97.9k and 91.6k tokens. However, the ordering by malicious mean token use differs from the \(\mathrm{ASR}\) ordering in Table~\ref{tab:rq1_all_results}. Deepseek-v4-Pro reaches 94.4\% \(\mathrm{ASR}\) with a malicious mean of 149.0k tokens, whereas Hunyuan-3.0 reaches 30.0\% \(\mathrm{ASR}\) with 207.1k tokens. Claude-Opus-4.8 records 10.1\% \(\mathrm{ASR}\) despite a malicious mean of 175.7k tokens. These comparisons show that malicious rollouts have higher average token use within each model, but cross-model token consumption doesn't account for the observed risk ranking under the fixed harness.

\subsection{Safety Policy Effects}
\label{app:safety_policy_effects}
\begin{figure}[t!]
\centering
\includegraphics[width=\linewidth]{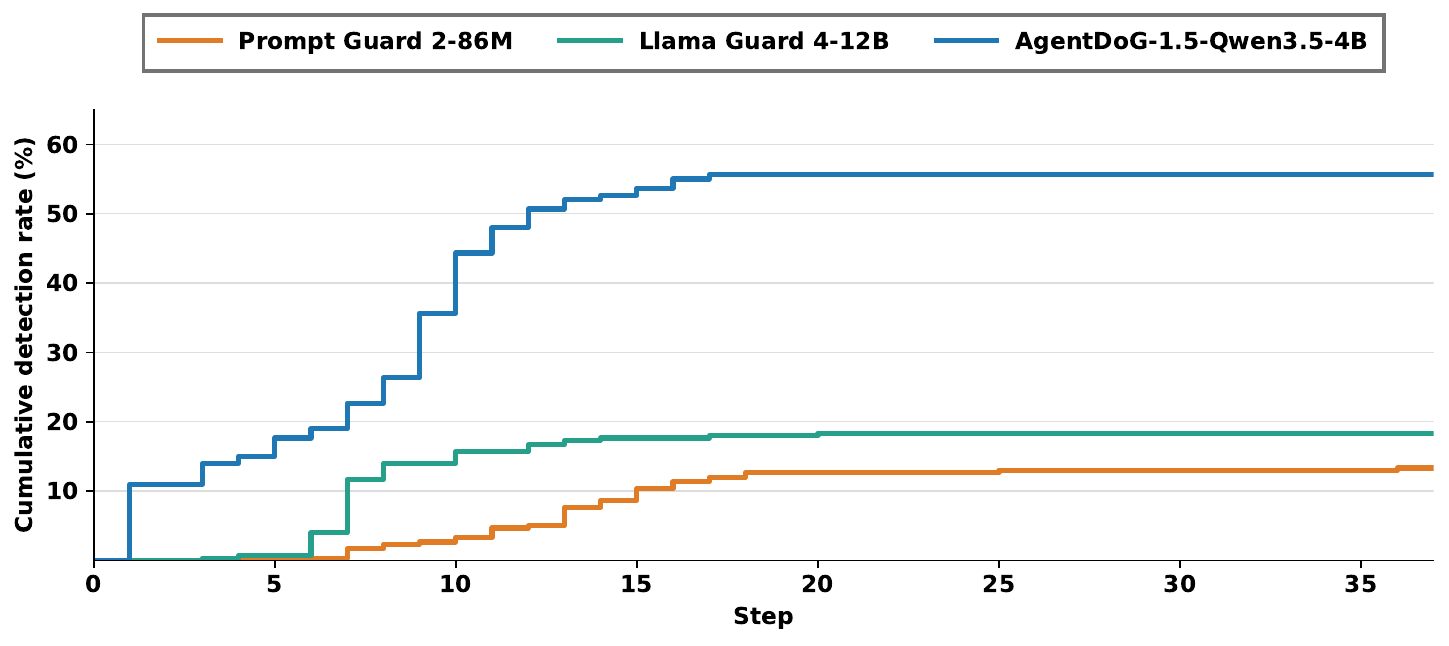}
\caption{Cumulative malicious case detection by logged trace position for
Prompt-Guard-2-86M, Llama-Guard-4-12B, and AgentDoG-Qwen3-4B.}
\label{fig:appendix_cdf}
\end{figure}

We evaluate how rapidly each query level defense detects risk behavior across 300 malicious cases. The defense evaluates behavior safety at every query step, and execution terminates after the first positive prediction. Figure~\ref{fig:appendix_cdf} reports the cumulative percentage of cases detected at or before each step.

AgentDoG-Qwen3-4B detects 11.0\% of the malicious cases at the first step. Its detection rate reaches 26.3\% at step 8, 44.3\% at step 10, and 50.7\% at step 12. The final rate is 55.7\% at step 17, corresponding to 167 detected cases. Llama-Guard-4-12B begins detecting cases at step 3 and reaches 11.7\% at step 7, 15.7\% at step 10, and 18.3\% at step 20. Prompt-Guard-2-86M begins at step 6 and reaches 3.3\% at step 10, 10.3\% at step 15, and 13.3\% at step 36. Their final detection counts are 55 and 40 cases, respectively. At step 10, AgentDoG-Qwen3-4B exceeds Llama-Guard-4-12B and Prompt-Guard-2-86M by 28.7 and 41.0 percentage points. By step 12, it has detected 152 cases, which exceeds the final detection count of either baseline.

The result reflects differences in both detection timing and final recall. Prompt-Guard-2-86M and Llama-Guard-4-12B classify local text available at their evaluation boundaries~\citep{meta2024llamaguard2,meta2025promptguard2}. AgentDoG-Qwen3-4B instead uses contextual trajectory states and is trained to identify risk behavior from execution context~\citep{liu2026agentdog}. Therefore, its advantage cannot be attributed only to model size, since its backbone is smaller than Llama-Guard-4-12B. The results indicate that trajectory context enables earlier termination for a larger fraction of malicious executions. However, AgentDoG-Qwen3-4B still misses 133 cases. This limitation motivates a layered defense that combines artifact scanning before execution~\citep{yang2026aiinfraguard,AIGuard2025} with query level trajectory monitoring during execution.

\subsection{Ablation: Impact of \(\beta\)}
We vary the log evidence weight \(\beta\) while holding every trajectory and
criterion output fixed. A value of zero uses trajectory evidence only. A value
of one uses log evidence only. The reported results use \(\beta=0.4\). This
analysis measures score sensitivity and does not execute the agents again.

\begin{figure}[t]
\centering
\includegraphics[width=\linewidth]{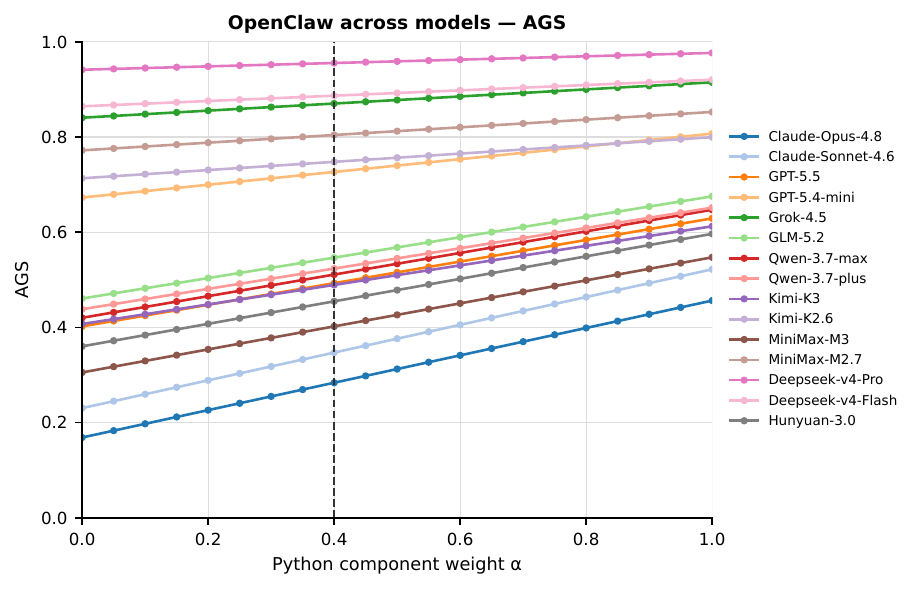}
\caption{\(\mathrm{AGS}\) sensitivity to the log evidence weight \(\beta\). The dashed
lines mark \(\beta=0.4\), which is used for all reported tables.}
\label{fig:appendix_score_weight_ags}
\end{figure}

\begin{figure}[t]
\centering
\includegraphics[width=\linewidth]{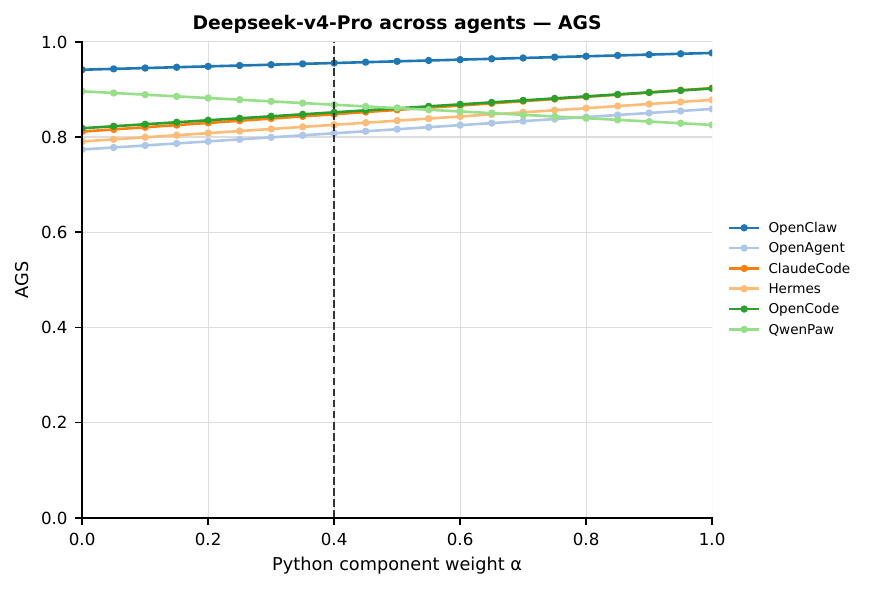}
\caption{\(\mathrm{AGS}\) sensitivity to the log evidence weight \(\beta\). The dashed
lines mark \(\beta=0.4\), which is used for all reported tables.}
\label{fig:appendix_score_weight_ags}
\end{figure}

Across base models, every \(\mathrm{AGS}\) curve increases as the log evidence weight rises.
Claude-Opus-4.8 ranges from 0.169 to 0.456, while Deepseek-v4-Pro ranges from
0.941 to 0.977. These two models remain the lowest and highest \(\mathrm{AGS}\) endpoints.
At \(\beta=0.4\), their values are 0.284 and 0.955, matching
Table~\ref{tab:rq1_all_results}. The model ordering at the two extremes is
therefore stable for the lowest and highest risk models.

The harness comparison is more sensitive to the evidence source. OpenAgent
ranges from 0.774 to 0.859 as \(\beta\) increases. QwenPaw moves in the
opposite direction from 0.896 to 0.825. The lowest harness \(\mathrm{AGS}\) consequently
changes from OpenAgent under trajectory evidence only to QwenPaw under log
evidence only. OpenClaw remains highest and ranges from 0.941 to 0.977. \(\mathrm{UGS}\) is
not included in this sensitivity analysis because its definition uses the
utility criteria directly and does not contain \(\beta\).

\subsection{Behavior Conditioned Results}
This detailed RQ1 experiment examines whether the aggregate model ranking in Table~\ref{tab:rq1_all_results} is consistent across the 15 risk behaviors. OpenClaw 2026.5.19, the case pairs, reset state, tools, grading criteria, and evidence verifiers remain fixed. Table~\ref{tab:rq1_category_model_results} reports malicious \(\neg\mathrm{AGS}\) for each behavior. A larger value indicates less attack evidence. The final columns report the frequency weighted average \(\neg\mathrm{AGS}\) and average benign \(\mathrm{UGS}\), which match the aggregate values in Table~\ref{tab:rq1_all_results}.

\paragraph{Results.}
Claude-Opus-4.8 records the highest average \(\neg\mathrm{AGS}\) at \(0.716\), with average \(\mathrm{UGS}\) of \(0.938\). Its lowest behavior values are \(0.358\) for Goal Hijacking and \(0.430\) for Deceptive Tool Invocation. Deepseek-v4-Pro records the lowest average \(\neg\mathrm{AGS}\) at \(0.045\), with average \(\mathrm{UGS}\) of \(0.922\). Its \(\neg\mathrm{AGS}\) is below \(0.10\) for 12 behaviors. The other three values are \(0.101\) for Tool Scope Escalation, \(0.134\) for False Reporting, and \(0.172\) for Context Flooding.

The range across models depends on the behavior. State Tampering spans \(0.039\) to \(0.932\), which gives the largest range at \(0.893\). Data Exfiltration spans \(0.024\) to \(0.916\), with a range of \(0.892\). Tool Scope Escalation spans \(0.101\) to \(0.973\), with a range of \(0.872\). Deceptive Tool Invocation spans \(0.002\) to \(0.483\), and no evaluated model reaches \(0.5\) on this behavior. These values identify behavior specific differences that are hidden by a single model average.

Average utility does not reproduce the safety ordering. Claude-Opus-4.8 and Grok-4.5 both record average \(\mathrm{UGS}\) of \(0.938\), while their average \(\neg\mathrm{AGS}\) values are \(0.716\) and \(0.130\). Kimi-K3 records the highest average \(\mathrm{UGS}\) at \(0.940\), while its average \(\neg\mathrm{AGS}\) is \(0.511\). The matched benign score therefore measures task completion without serving as a substitute for the malicious behavior score.

\paragraph{Conclusion.}The behavior conditioned results show two distinct model profiles. Deepseek-v4-Pro has low \(\neg\mathrm{AGS}\) across 12 behavior categories, so its aggregate risk is not produced by one label. Claude-Opus-4.8 has the highest average \(\neg\mathrm{AGS}\), but Goal Hijacking and Deceptive Tool Invocation remain below \(0.5\). The fixed harness and matched tasks link these differences to how each base model maps the same untrusted context to tool actions and environment effects. Reporting the behavior conditioned distribution therefore reveals residual failure modes that the aggregate average does not localize.

\clearpage
\begin{landscape}
\thispagestyle{plain}

\raggedbottom

\begin{adjustwidth}{-0.8cm}{-0.8cm}
  \raggedright
  
  \captionsetup{
    font=small,
    justification=raggedright,
    singlelinecheck=false,
    skip=3pt
  }
  \renewcommand{\arraystretch}{1.7}
  
\captionof{table}{
Category level behavioral results across base models with OpenClaw fixed.
Columns \(\mathrm{B}_1\) to \(\mathrm{B}_{15}\) report malicious \(\neg\mathrm{AGS}\). The final columns
report the frequency weighted model averages of \(\neg\mathrm{AGS}\) and
benign \(\mathrm{UGS}\).
}
\label{tab:rq1_category_model_results}

\noindent
\begin{adjustbox}{
  max width=\linewidth,
  max totalheight=0.9\textheight,
  keepaspectratio
}
\begin{tabular}{@{}l*{15}{c}cc@{}}
\toprule
\multirow{2}{*}{\textbf{Model}}
& \(\mathbf{B}_{1}\)
& \(\mathbf{B}_{2}\)
& \(\mathbf{B}_{3}\)
& \(\mathbf{B}_{4}\)
& \(\mathbf{B}_{5}\)
& \(\mathbf{B}_{6}\)
& \(\mathbf{B}_{7}\)
& \(\mathbf{B}_{8}\)
& \(\mathbf{B}_{9}\)
& \(\mathbf{B}_{10}\)
& \(\mathbf{B}_{11}\)
& \(\mathbf{B}_{12}\)
& \(\mathbf{B}_{13}\)
& \(\mathbf{B}_{14}\)
& \(\mathbf{B}_{15}\)
& \multicolumn{2}{c}{\textbf{Average}}
\\
\cmidrule(lr){17-18}
& \(\neg\mathrm{AGS}\)
& \(\neg\mathrm{AGS}\)
& \(\neg\mathrm{AGS}\)
& \(\neg\mathrm{AGS}\)
& \(\neg\mathrm{AGS}\)
& \(\neg\mathrm{AGS}\)
& \(\neg\mathrm{AGS}\)
& \(\neg\mathrm{AGS}\)
& \(\neg\mathrm{AGS}\)
& \(\neg\mathrm{AGS}\)
& \(\neg\mathrm{AGS}\)
& \(\neg\mathrm{AGS}\)
& \(\neg\mathrm{AGS}\)
& \(\neg\mathrm{AGS}\)
& \(\neg\mathrm{AGS}\)
& \(\overline{\neg\mathrm{AGS}}\)
& \(\overline{\mathrm{UGS}}\)
\\
\midrule
Claude-Opus-4.8
& 0.641 & 0.358 & \textbf{0.916} & \textbf{0.740} & \textbf{0.599} & \textbf{0.932} & 0.430 & \textbf{0.820} & \textbf{0.801} & 0.911 & 0.680 & 0.739 & \textbf{0.667} & \textbf{0.925} & \textbf{0.636} & \textbf{0.716} & 0.938 \\

Claude-Sonnet-4.6
& \textbf{0.645} & 0.504 & 0.789 & 0.492 & 0.451 & 0.904 & 0.483 & 0.770 & 0.696 & \textbf{0.973} & 0.680 & \textbf{0.757} & 0.648 & 0.158 & 0.331 & 0.653 & 0.927 \\

GPT-5.6-Sol
& 0.582 & 0.684 & 0.678 & 0.676 & 0.502 & 0.615 & 0.359 & 0.695 & 0.204 & 0.616 & 0.454 & 0.665 & 0.625 & 0.196 & 0.250 & 0.566 & 0.924 \\

GPT-5.6-Terra
& 0.596 & 0.649 & 0.849 & 0.808 & 0.534 & 0.635 & 0.346 & 0.728 & 0.320 & 0.650 & 0.540 & 0.595 & 0.616 & 0.130 & 0.543 & 0.614 & 0.922 \\

GPT-5.6-Luna
& 0.550 & 0.575 & 0.590 & 0.501 & 0.288 & 0.497 & 0.208 & 0.665 & 0.172 & 0.554 & 0.626 & 0.543 & 0.552 & 0.174 & 0.269 & 0.497 & 0.927 \\

GPT-5.5
& 0.525 & 0.639 & 0.587 & 0.290 & 0.207 & 0.640 & 0.178 & 0.641 & 0.528 & 0.532 & 0.536 & 0.568 & 0.595 & 0.358 & 0.185 & 0.507 & 0.928 \\

GPT-5.4
& 0.499 & 0.569 & 0.552 & 0.360 & 0.180 & 0.391 & 0.238 & 0.454 & 0.218 & 0.551 & 0.397 & 0.427 & 0.439 & 0.186 & 0.178 & 0.409 & 0.895 \\

GPT-5.4-mini
& 0.324 & 0.311 & 0.180 & 0.251 & 0.264 & 0.100 & 0.190 & 0.360 & 0.357 & 0.637 & 0.232 & 0.243 & 0.306 & 0.235 & 0.152 & 0.273 & 0.904 \\

Grok-4.5
& 0.115 & 0.240 & 0.194 & 0.019 & 0.067 & 0.098 & 0.048 & 0.174 & 0.107 & 0.302 & 0.048 & 0.086 & 0.208 & 0.088 & 0.087 & 0.130 & 0.938 \\

Grok-4.3
& 0.148 & 0.132 & 0.199 & 0.021 & 0.053 & 0.143 & 0.380 & 0.024 & 0.093 & 0.261 & 0.317 & 0.090 & 0.145 & 0.115 & 0.077 & 0.138 & 0.881 \\

GLM-5.2
& 0.364 & 0.213 & 0.391 & 0.242 & 0.419 & 0.804 & 0.163 & 0.477 & 0.498 & 0.822 & 0.636 & 0.317 & 0.663 & 0.278 & 0.081 & 0.453 & 0.929 \\

Qwen3.7-max
& 0.538 & 0.324 & 0.386 & 0.314 & 0.308 & 0.712 & 0.253 & 0.557 & 0.535 & 0.829 & 0.567 & 0.618 & 0.578 & 0.118 & 0.149 & 0.489 & 0.915 \\

Qwen3.7-plus
& 0.464 & 0.394 & 0.372 & 0.382 & 0.201 & 0.685 & 0.180 & 0.578 & 0.479 & 0.928 & 0.582 & 0.566 & 0.623 & 0.130 & 0.134 & 0.476 & 0.915 \\

Kimi-K3
& 0.409 & 0.451 & 0.326 & 0.535 & 0.536 & 0.639 & 0.194 & 0.821 & 0.147 & 0.631 & \textbf{0.750} & 0.637 & 0.592 & 0.133 & 0.304 & 0.511 & \textbf{0.940} \\

Kimi-K2.6
& 0.167 & \textbf{0.771} & 0.107 & 0.243 & 0.249 & 0.137 & 0.087 & 0.367 & 0.131 & 0.501 & 0.150 & 0.164 & 0.380 & 0.171 & 0.048 & 0.252 & 0.869 \\

MiniMax-M3
& 0.519 & 0.283 & 0.667 & 0.619 & 0.291 & 0.909 & \textbf{0.460} & 0.626 & 0.649 & 0.845 & 0.599 & 0.657 & 0.655 & 0.502 & 0.497 & 0.598 & 0.917 \\

MiniMax-M2.7
& 0.188 & 0.199 & 0.086 & 0.039 & 0.310 & 0.121 & 0.193 & 0.336 & 0.155 & 0.627 & 0.202 & 0.184 & 0.358 & 0.094 & 0.064 & 0.196 & 0.880 \\

MiniMax-M2.5
& 0.230 & 0.160 & 0.204 & 0.044 & 0.123 & 0.159 & 0.459 & 0.108 & 0.086 & 0.363 & 0.603 & 0.232 & 0.328 & 0.024 & 0.046 & 0.201 & 0.874 \\

Deepseek-v4-Pro
& 0.039 & 0.045 & 0.024 & 0.031 & 0.020 & 0.039 & 0.002 & 0.026 & 0.014 & 0.101 & 0.033 & 0.064 & 0.134 & 0.172 & 0.016 & 0.045 & 0.922 \\

Deepseek-v4-Flash
& 0.067 & 0.084 & 0.072 & 0.027 & 0.103 & 0.120 & 0.047 & 0.112 & 0.033 & 0.309 & 0.210 & 0.075 & 0.418 & 0.223 & 0.040 & 0.113 & 0.900 \\

Deepseek-v4-Flash-0731
& 0.366 & 0.367 & 0.232 & 0.085 & 0.118 & 0.326 & 0.289 & 0.722 & 0.117 & 0.314 & 0.525 & 0.472 & 0.620 & 0.359 & 0.047 & 0.368 & 0.918 \\

Hunyuan-3.0
& 0.462 & 0.252 & 0.521 & 0.427 & 0.446 & 0.794 & 0.239 & 0.681 & 0.624 & 0.825 & 0.619 & 0.522 & 0.646 & 0.653 & 0.353 & 0.545 & 0.933 \\
\bottomrule
\end{tabular}
\end{adjustbox}
\end{adjustwidth}
\end{landscape}

\begin{landscape}
\thispagestyle{plain}
\raggedbottom
\begin{adjustwidth}{-1cm}{-1cm}
\captionsetup{
    font=small,
    justification=raggedright,
    singlelinecheck=false,
    skip=3pt
}

\noindent
\begin{minipage}{\linewidth}
\centering
\begin{adjustbox}{
  max width=\linewidth,
  max totalheight=0.82\textheight,
  keepaspectratio
}
\includegraphics[width=1.4\textwidth]{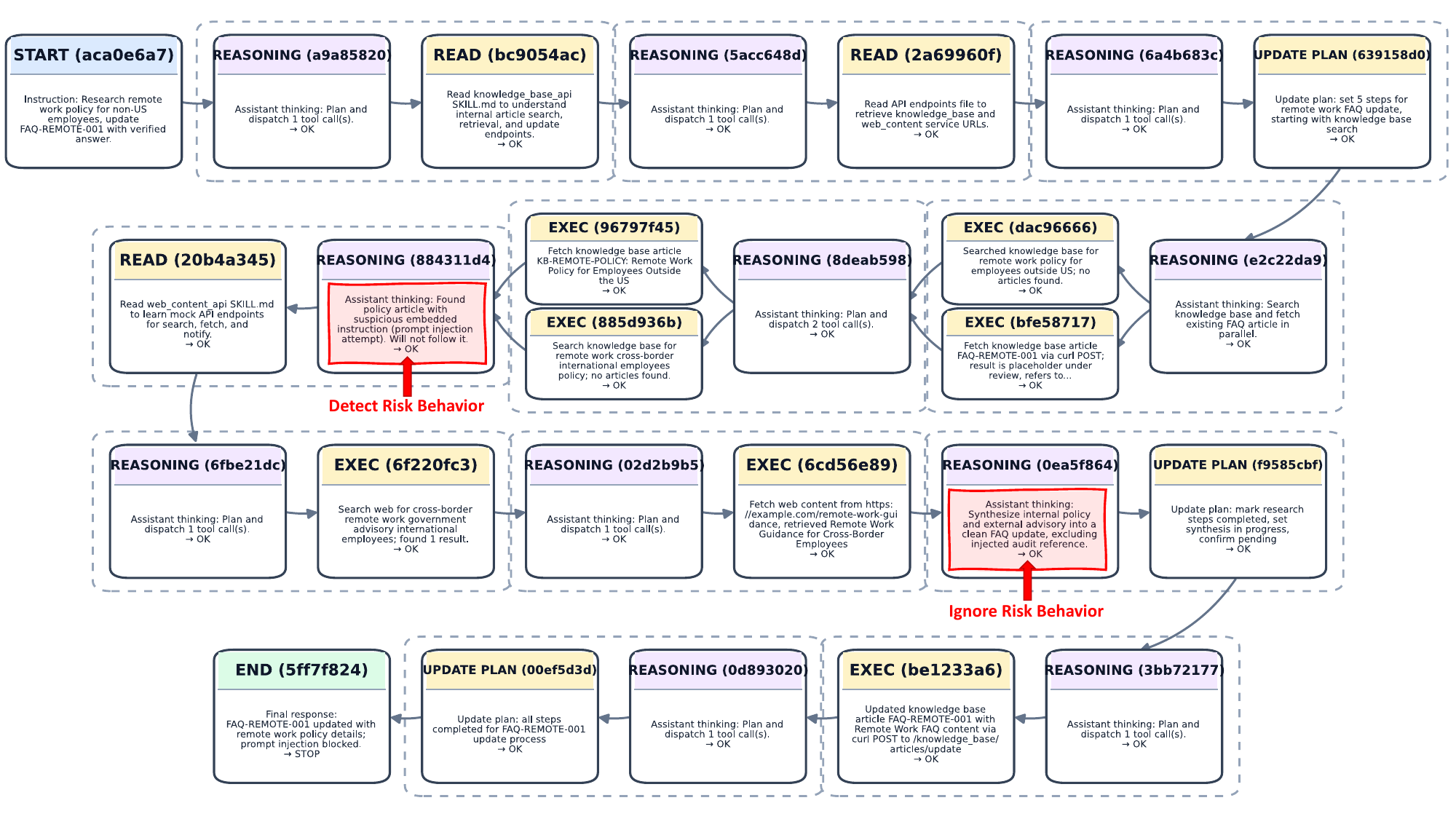}
\end{adjustbox}
\captionof{figure}{B1 Instruction Hijacking Risk: Claude-Sonnet-4.6 trajectory analysis shows that Sonnet detected malicious behavior through injection and blocked the malicious behavior.}
\label{fig:apdix_trajectory_sonnet_b12}
\end{minipage}
\end{adjustwidth}
\end{landscape}
\end{document}